\documentclass[preprint,aps,prd,showpacs,superscriptaddress,nofootinbib,tightenlines]{revtex4}
\usepackage{amsfonts}
\usepackage{multirow}
\usepackage{mathrsfs}
\usepackage{graphicx}
\usepackage{amsmath}
\usepackage{amssymb}
\usepackage{bm}
\usepackage{bbm}
\usepackage{slashbox}

\def\OMIT#1{}

\newcommand{\nn}{\nonumber}

\newcommand{\beq}{\begin{equation}}
\newcommand{\eeq}{\end{equation}}
\newcommand{\bqa}{\begin{eqnarray}}
\newcommand{\eqa}{\end{eqnarray}}

\begin{document}
%\preprint{}
%%%%%%%%%%%%%%%%%%%%%%%%%%%%%%%%%%%%%%%%%%%%%%%%%%%%%%%%%%%%%%%%%%%%%%%%%%%%%%
\title{\mbox{}\\[10pt]
Inclusive $\bm{h}_{\bm c}$ production at $\bm{B}$ factories}
%%%%%%%%%%%%%%%%%%%%%%%%%%%%%%%%%%%%%%%%%%%%%%%%%%%%%%%%%%%%%%%%%%%%%%%%%%%%%%

\author{Yu Jia\footnote{jiay@ihep.ac.cn}}
\affiliation{Institute of High Energy Physics, Chinese Academy of
Sciences, Beijing 100049, China\vspace{0.2cm}}
\affiliation{Theoretical Physics Center for Science Facilities,
Chinese Academy of Sciences, Beijing 100049, China\vspace{0.2cm}}

\author{Wen-Long Sang\footnote{wlsang@ihep.ac.cn}}
\affiliation{Institute of High Energy Physics, Chinese Academy of
Sciences, Beijing 100049, China\vspace{0.2cm}}
\affiliation{Theoretical Physics Center for Science Facilities,
Chinese Academy of Sciences, Beijing 100049, China\vspace{0.2cm}}

\author{Jia Xu\footnote{xuj@ihep.ac.cn}} \affiliation{Institute of High
Energy Physics, Chinese Academy of Sciences, Beijing 100049,
China\vspace{0.2cm}}

\date{\today}
%%%%%%%%%%%%%%%%%%%%%%%%%%%%%%%%%%%%%%%%%%%%%%%%%%%%%%%%%%%%%%%%%%%%%%%%%%%%%%
\begin{abstract}
Within the nonrelativistic QCD (NRQCD) factorization framework, we
investigate the inclusive production of the $h_c$ meson associated
with either light hadrons or charmed hadrons at $B$ factory energy
$\sqrt{s}=10.58$ GeV. Both the leading color-singlet and color-octet
channels are included. For the $h_c$ production associated with
light hadrons, the total production rate is dominated by the
color-octet channel, thus the future measurement of this process may
impose useful constraint on the value of the color-octet matrix
element $\langle {\cal O}^{h_c}_8({}^1S_0)\rangle$; for the $h_c$
production associated with charmed hadrons, the total production
rate is about one order of magnitude smaller, and dominated by the
color-singlet channel.
\end{abstract}

%%%%%%%%%%%%%%%%%%%%%%%%%%%%%%%%%%%%%%%%%%%%%%%%%%%%%%%%%%%%%%%%%%%%%%%%%%%%%%

%11.10.Hi Renormalization group evolution of parameters
%12.38.Bx Perturbative calculations
%12.38.Cy Summation of perturbation theory
%12.38.-t Quantum chromodynamics
%12.39.St Factorization
%13.20.Gd Decays of J/¦×, ¦´, and other quarkonia
%13.60.Le Meson production
%13.60.Hb Total and inclusive cross sections (including deep-inelastic processes)
%13.87.Fh Fragmentation into hadrons
%14.40.Pq Heavy quarkonia
%%%%%%%%%%%%%%%%%%%%%%%%%%%%%%%%%%%%%%%%%%%%%%%%%%%%%%%%%%%%%%%%%%%%%%%%%%%%%%

\maketitle

\section{Introduction}

The lowest-lying $c\bar{c}({}^1P_1)$ state, the $h_c(1P)$ meson, is
the last one found among all the charmonium members below open charm
threshold. This elusive particle was not firmly established until
2005, through the isospin-violating decay $\psi'\to \pi^0(\to
\gamma\gamma)h_c(\to \eta_c\gamma)$ by CLEO
Collaboration~\cite{Rosner:2005ry}, as well as through the process
$p\bar{p}\to h_c\to \eta_c\gamma$ by the E835
experiment~\cite{Andreotti:2005vu}. Quite recently, the analogous
$^1P_1$ members in the bottomonium family, the $h_b(1P, 2P)$ mesons,
have also been observed by the \textsc{Belle} Collaboration through
the process $e^+e^-\to \Upsilon(5S)\to
h_b(nP)+\pi^+\pi^-$~\cite{Adachi:2011ji}.

The quite accurate measurements of the mass of the $h_c$ (also
$h_b(1P, 2P)$) implies a rather small $P$-wave hyperfine mass
splitting. This is theoretically intriguing since it might be able
to impose some severe constraint on the quark spin-spin interaction
as well as possible charm meson loop effect.

Aside from its mass~\cite{Dobbs:2008ec,Ablikim:2010rc}, our
knowledge about the $h_c$ state is still quite limited. The $h_c$
appears to be a narrow resonance with the total width of $0.73\pm
0.45\pm 0.28$ MeV~\cite{Ablikim:2010rc}. So far, only two decay
channels of the $h_c$ have been measured, one is the hadronic decay
$h_c\to 2(\pi^+ \pi^-)\pi^0$~\cite{Adams:2009aa}, and the other is
the much more abundant $E1$ transition $h_c\to \eta_c\gamma$.
Recently BES~III experiment has measured the absolute branching
fraction of the latter process and gives ${\cal B}(h_c\to \gamma
\eta_c)=(54.3\pm 6.7\pm5.2)\%$~\cite{Ablikim:2010rc}.

In contrast to the decay, our understanding of the $h_c$ production
is even poorer. The only measurement of the $h_c$ production is from
a recent \textsc{CLEO} experiment, by observing the process
$e^+e^-\to h_c \pi^+\pi^-$ at $\sqrt{s}=4.170$
GeV~\cite{CLEO:2011aa}. On the theoretical side, rather few works on
$h_c$ production are scattered in literature. Among these studies,
are $h_c$ production in $B$ meson inclusive
decay~\cite{Bodwin:1992qr,Beneke:1998ks}, $h_c$
photoproduction~\cite{Fleming:1998md}, $h_c$
hadroproduction~\cite{Sridhar:2008sc,Qiao:2009zg}.  It is
interesting to compare this situation with highly intensive studies
on $J/\psi$ production in various collision
experiments~\cite{Brambilla:2010cs}.

Our goal in this work is to carry out a detailed study on inclusive
$h_c$ production in $e^+e^-$ annihilation, which is specifically
relevant to the $B$ factory experiments. Specifically, we
investigate the inclusive production of the $h_c$ meson associated
both with light hadrons and with charmed hadrons. There are both
theoretical and experimental merits to study these processes. On the
theoretical side, since these processes are much simpler than the
$h_c$ production in hadronic collision, one expects to obtain more
precise prediction with less contamination from the nonperturbative
side of QCD. On the experimental side, since $e^+e^-$ collision
experiment possesses much clean background than the hadronic
collider, and the $B$ factories have already accumulated a quite
large data sample near the $\Upsilon(4S)$ resonance, it could well
be the most likely place to unambiguously observe the elusive $h_c$
signal.

Our study is based on the nonrelativistic QCD (NRQCD) factorization
approach, a widely-accepted framework to deal with inclusive
quarkonium production~\cite{Bodwin:1994jh}, which heavily exploits
the nonrelativistic nature of heavy quarkonium.  A highlight of this
approach is the so-called color-octet mechanism, which serves as an
indispensable ingredient in order to give a meaningful prediction
for the $P$ wave quarkonium production~\cite{Bodwin:1992qr}.

As we will see, for the process $e^+e^-\to h_c+$ light hadrons, the
color-octet mechanism indeed plays a pivotal role in rendering
infrared-finite prediction for the inclusive $h_c$ production rate.
Furthermore, this process is found to be dominated by the
color-octet channel. Our study reveals that this process may have a
sizable total cross section that is comparable in magnitude with
that of $e^+e^-\to J/\psi +$ light hadrons, which has been measured
some time ago by the $B$ factory experiments~\cite{Pakhlov:2009nj}.
Future measurement of this process at the $B$ factories may provide
an explicit test on the color-octet mechanism, as well as put some
useful constraint on the value of the color-octet matrix element for
$h_c$~\footnote{The majority of the results in this paper has
already been presented in Ref.~\cite{Xu:2011:thesis}.}.

The rest of the paper is organized as follows.
%--------------------------------
In Sec.~\ref{NRQCD:factorization:matching}, we present the NRQCD
factorization formulas for $h_c$ inclusive production in $e^+e^-$
annihilation, accurate at lowest order in $v$ (the characteristic
velocity of charm quark inside the $h_c$ meson).
%--------------------------------
In Sec.~\ref{matching:hc:gg}, by employing the perturbative matching
ansatz, we determine the infrared-finite color-singlet and
color-octet short-distance coefficients associated with the process
$e^+e^-\to h_c+$ light hadrons.
%--------------------------------
In Sec.~\ref{matching:hc:ccbar}, we determine the color-singlet and
color-octet short-distance coefficients associated with the process
$e^+e^-\to h_c+$ charmed hadrons.
%--------------------------------
Based on our calculation in previous sections, we devote
Sec.~\ref{phenomenology} to exploring the observation prospects of
$h_c$ production at $B$ factories.
%--------------------------------
Finally we summarize in Sec.~\ref{summary}.
%--------------------------------
In Appendix~\ref{appendix:DR:isolate:IR:div}, we explain some
technical details in isolating the infrared divergence for
$e^+e^-\to h_c+gg$ in dimensional regularization.
%--------------------------------
In Appendix~\ref{appendix:etac:c:cb:dis:frag}, we present the
color-singlet short-distance coefficient for the process $e^+e^-\to
\eta_c+c\bar{c}$.

%--------------------------------------------------------------------
\section{NRQCD factorization formula for ${\bm h}_{\bm c}$ production}
\label{NRQCD:factorization:matching}
%--------------------------------------------------------------------

NRQCD factorization formalism is a systematic tool for analyzing the
inclusive production of heavy quarkonium~\cite{Bodwin:1994jh}. The
production rate can be expressed as a sum of products of
short-distance coefficients and nonperturbative, albeit universal
vacuum NRQCD matrix elements, whose importance is organized by the
typical quark velocity, $v$. In this work, we will consider
$e^+e^-\rightarrow h_c+X$ in this factorization framework. At the
lowest order in $v$, the velocity counting rule implies that the
cross section of $h_c$ has the following form:
%--------------------------
\bqa
%--------------------------
d\sigma[e^+e^-\rightarrow h_c+X] = {dF_1 \over m_c^4} \langle
\mathcal{O}_1^{h_c}(^1P_1) \rangle + {dF_8 \over m_c^2} \langle
\mathcal{O}_8^{h_c}(^1S_0) \rangle,
%--------------------------
\label{NRQCD:factorization:formula}
%--------------------------
\eqa
%--------------------------
where the color-singlet operator $\mathcal{O}_1^{h_c}(^1P_1)$ and
the color-octet operator $\mathcal{O}_8^{h_c}(^1S_0)$ have been
introduced in \cite{Bodwin:1994jh}. It is interesting to contrast
this $P$-wave quarkonium production process with the $S$-wave onium
production, where the color-octet operator matrix element first
comes into play only at relative $O(v^4)$.

$dF_1 $, $dF_8$ are infrared-finite short-distance coefficients
associated with the vacuum matrix elements of color-singlet and
-octet NRQCD production operators. Since they are insensitive to the
long-distance strong interaction dynamics, a standard way of
determining them is through the perturbative matching procedure:
replacing the $h_c$ state appearing in
(\ref{NRQCD:factorization:formula}) with the free on-shell
$c\bar{c}(n)$ states ($n={}^1S_0^{(8)}$ or ${}^1P_1^{(1)}$), and
computing both sides of (\ref{NRQCD:factorization:formula}) using
perturbative QCD and perturbative NRQCD, respectively, then
enforcing that they generate identical results. Finally, one then
solves two linear equations to identify the two unknown
coefficients, order by order in $\alpha_s$~\footnote{In this work,
we are only looking for the $h_c$ energy distribution and total
cross section, rather than its angular distribution. To this
purpose, we may adopt a standard shortcut to simplify the
intermediate calculations~\cite{Keung:1980ev}. First compute the
virtual photon decay into $h_c$, then use the following formula to
convert the decay rate into $h_c$ cross section: $d \sigma[e^+
e^-\to h_c(P)+X]/d P^0 = {4\pi\alpha\over s^{3/2}}\,d
\Gamma[\gamma^\ast\to h_c(P)+X]/d P^0$, where $P^\mu$ represents the
4-momentum of the $h_c$ in the $e^+e^-$ center-of-mass frame.
\label{conversion:decay:production}}.

The perturbative matching for $e^+e^-\to h_c+$ charmed hadrons is
straightforward. In contrast, the matching procedure for $e^+e^-\to
h_c+$ light hadrons is more subtle and involved, since the QCD side
calculation for the color-singlet channel $d\sigma[e^+e^-\rightarrow
c\bar{c}({}^1P_1^{(1)})+gg]$ contains infrared divergence, which
must be absorbed into the color-octet matrix element to render an
infrared (IR) finite color-singlet short-distance coefficient.

%--------------------------------------------------------------------
\section{${\bm h}_{\bm c}$ production associated with light hadrons}
\label{matching:hc:gg}
%--------------------------------------------------------------------

In this section, we will apply the perturbative matching procedure
to deduce the short-distance coefficients for inclusive $h_c$
production associated with light hadrons at $B$ factories, i.e.,
$e^+e^-\to h_c+{\rm light\;hadrons}$. For clarity, we will always
attach a superscript ``LH" to both $F_1$ and $F_8$ in this section,
to differentiate them from the analogous coefficients for the $h_c$
production associated with charmed hadrons, which will be reported
in the next section.

%--------------------------------------------------------------------
\subsection{Determining $d F_8^{\rm LH}$}
\label{match:hc:gg:dF8}
%--------------------------------------------------------------------

We begin with calculating the differential short-distance
coefficient $d F_8^{\rm LH}$ affiliated with the color-octet
operator. At the lowest order in $\alpha_s$, only two Feynman
diagrams need to be considered for the process $e^+e^-\to
c\bar{c}({}^1S_0^{(8)})+g$.

As mentioned before, the differential cross section
$d\sigma[e^+e^-\rightarrow c\bar{c}({}^1P_1)+gg]/dz$ would develop
an IR divergence as one of the gluons gets soft. In order to deduce
the IR finite color-singlet coefficient $d F_1^{\rm LH}$, the
perturbative factorization formula
(\ref{NRQCD:factorization:formula}) implies that we should also
consider the color-octet coefficient $dF_8$ multiplied by the
renormalized $O(\alpha_s)$ color-octet matrix element, which is
generally IR divergent. Throughout this work we find it most
convenient to employ the dimensional regularization (DR) to
regularize both UV and IR divergences. Therefore, it is necessary to
compute $e^+e^-\to c\bar{c}({}^1S_0^{(8)})+g$ in $D=4-2\epsilon$
spacetime dimensions.

It is convenient to use the covariant spin projection
method~\cite{Petrelli:1997ge} to compute the amplitude of
$\gamma^*\to c\bar{c}({}^1S_0^{(8)})+g$ in $D$ spacetime dimensions.
A subtlety is that the appearance of $\gamma_5$ from the
spin-singlet projector, which requires some care to handle it in $D$
dimensions. For consistency, we adopt the 't Hooft-Veltman (HV)
prescription~\cite{'tHooft:1972fi}, and utilize West's formula to
calculate the trace involving one $\gamma_5$ and a string of Dirac
matrices~\cite{West:1991xv}. The Levi-Civita tensor is assumed as a
4-dimensional object~\footnote{As a crosscheck, we have also tried
to treat the Levi-Civita tensor as a $D$-dimensional object when
computing the squared amplitude. After matching is done, one is
justified to return to $4$ dimensions. It turns out that this
alternative prescription leads to the identical short-distance
coefficients $dF_8$ and $dF_1$, as given in
(\ref{color-octet:short-distance:coef:4:dimensions}) and
(\ref{differential:F1:L.H.}) in the text.}.

We will always work in the $e^+e^-$ center-of-mass frame throughout
this paper, where $\sqrt{s}$ denotes the $e^+e^-$ center-of-mass
energy.  For notational simplicity, we define the energy fraction of
$h_c$, $z \equiv 2 P^0/\sqrt{s}$, as well as the ratio of charm
quark mass over center-of-mass energy, $r \equiv 4m_c^2/s$. The
differential two-body phase space in $D$ dimensions can be expressed
as
%---------------------------------
\bqa
%---------------------------------
d\Phi_2= {c_\epsilon \over 8\pi}
s^{-\epsilon}(1-r)^{1-2\epsilon}\delta(1+r-z)dz,
%---------------------------------
\label{2-body phase space}
%---------------------------------
\eqa
%---------------------------------
where $c_\epsilon \equiv (4\pi)^\epsilon {\Gamma(1-\epsilon)\over
\Gamma(2-2\epsilon)}$.

Comparing both sides of (\ref{NRQCD:factorization:formula}) at
$O(\alpha_s)$, it is easy to find the differential coefficient
$dF_8^{\rm LH}$ in $D$ dimensions:
%---------------------------------
\bqa
%---------------------------------
{dF_8^{\rm LH} \over dz} &=& c_\epsilon\bigg
(\frac{\mu^2}{s}\bigg)^\epsilon {32\pi^2 e_c^2 \alpha^2\alpha_s m_c
\over 3 s^2} (1-r)^{1-2\epsilon} \delta(1+r-z),
%---------------------------------
\label{color-octet:short-distance:coef:D:dimensions}
%---------------------------------
\eqa
%---------------------------------
where $\mu$ is the compensating mass scale in DR, $e_c={2\over 3}$
is the electric charge of charm quark.

In the limit $\epsilon\to 0$, the differential color-octet
coefficient reduces to
%---------------------------------
\bqa
%---------------------------------
{dF_8^{\rm LH} \over dz} &=&  {32\pi^2 e_c^2 \alpha^2\alpha_s m_c
\over 3 s^2} (1-r)\,\delta(1+r-z).
%---------------------------------
\label{color-octet:short-distance:coef:4:dimensions}
%---------------------------------
\eqa
%---------------------------------

\subsection{Determining $dF_1^{\rm LH}$}

We proceed to determine the color-singlet coefficient for $h_c$
production associated with light hadrons in $e^+e^-$ annihilation,
$dF_1^{\rm LH}$. This can be expedited by replacing the $h_c$ state
with a free $c\bar{c}({}^1P_1^{(1)})$ pair, and matching both sides
of (\ref{NRQCD:factorization:formula}) that are computed in
perturbative QCD and perturbative NRQCD, respectively.

At the lowest order in $\alpha_s$, the color-singlet $h_c$
production associated with light hadrons can proceed through the
parton-level process $e^+e^-\to c\bar{c}(^1P^{(1)}_1)gg$, which can
be picturized  by six Feynman diagrams. Let $P$, $k_1$, $k_2$
signify the momenta of the $c\bar{c}(^1P^{(1)}_1)$ pair, gluon 1,
and gluon 2, respectively. We will always assume $P^2 \approx 4
m_c^2$. It is convenient to introduce three fractional energy
variables $z$, $x_1$ and $x_2$:
%---------------------------------
\bqa
%---------------------------------
z &=& {2P^0\over \sqrt{s}},\qquad x_1= {2k_1^0 \over
\sqrt{s}},\qquad x_2= {2k_1^0 \over \sqrt{s}},
%---------------------------------
\label{three:energy:fractions:def}
%---------------------------------
\eqa
%---------------------------------
which are subject to the constraint $x_1+x_2+z=2$, as required by
the energy conservation.

Since we employ the DR to regularize the potential IR divergences,
it is useful to write down the 3-body phase space integral in
$4-2\epsilon$ spacetime dimensions:
%--------------------------
\bqa
%--------------------------
\int d\Phi_3 &=& {c_\epsilon(4\pi)^{\epsilon} \over
\Gamma(1-\epsilon)} \bigg(\frac{s}{2}\bigg)^{1-2\epsilon} {1\over
(4\pi)^3} \int^{1+r}_{2\sqrt{r}} \! dz \int^{x_1^+}_{x_1^-} \!\!
dx_1 \, x_1^{-2\epsilon} (z^2-4 r)^{-\epsilon}
(1-\cos^2\!\theta)^{-\epsilon},
%--------------------------
\label{3-body:phase:space:D:dimensions}
%--------------------------
\eqa
%--------------------------
where $\theta$ signifies the angle between ${\bf P}$ and ${\bf k_1}$
in the $\gamma^*$ rest frame. For given $z$ and $x_1$, it can be
uniquely determined:
%--------------------------
\bqa
%--------------------------
\cos{\theta}=\frac{2(1+r-z) -x_1(2-z)}{x_1\sqrt{z^2-4r}}.
%--------------------------
\label{theta:def}
%--------------------------
\eqa
%--------------------------

Since we have taken the shortcut by first calculating the virtual
photon decay into 3-body final state, only two independent energy
fraction variables need be retained in the integration measure in
(\ref{3-body:phase:space:D:dimensions}). The integration limits of
$z$ have been explicitly labeled in
(\ref{3-body:phase:space:D:dimensions}), while the integration
boundaries of $x_1$, $x_1^{\pm}$, are parameterized as $a(z)\pm
b(z)$, where
%--------------------------
\begin{subequations}
%--------------------------
\bqa
%--------------------------
a(z)&=& {1 \over 2}(2-z),
%--------------------------
\\
%--------------------------
b(z)&=& {1 \over 2}\sqrt{z^2-4 r}.
%--------------------------
\eqa
%--------------------------
\label{def:a(z):b(z)}
%--------------------------
\end{subequations}
%--------------------------

We again use the spin projection technique~\cite{Petrelli:1997ge} to
compute the amplitude of $\gamma^*\to c\bar{c}({}^1P_1^{(1)})+gg$ in
$D$ dimensions and use the HV prescription to handle the trace
involving $\gamma_5$. After squaring the amplitude and summing over
polarizations and colors, we separate the squared amplitude into two
pieces:
%--------------------------
\bqa
%--------------------------
 && \sum_{\rm Pol,\,Col}
 \left|{\mathcal A}[\gamma^*\to c\bar{c}({}^1P_1^{(1)})+g g]\right|^2
  =
 I_{\rm div}(x_1,z)+I_{\rm
fin}(x_1,z),
%--------------------------
\label{ampl:squared:gamma*:ccbargg}
%--------------------------
 \eqa
%--------------------------
where
%--------------------------
\bqa
%--------------------------
 I_{{\rm div}}(x_1,z)
 &=& {2^{16} \pi^3  e_c^2
 \alpha C_F \alpha_s^2 \mu^{4\epsilon}\over s^2} (1-\epsilon)
 \left[ {1\over (1+r-z-x_1)^2}+ {1 \over (1+r-z-x_2)^2}\right]
%--------------------------
\label{Idiv:def}
%--------------------------
\eqa
%--------------------------
represents the term that would bring forth an IR singularity when
integrating over two distinct phase space corners: $z\to 1+r, x_1\to
0$ and $z\to 1+r, x_2\to 0$, where one of the gluons become soft.
The symbol $I_{{\rm fin}}$ denotes the remainder of the squared
amplitude which contains no terms as dangerous as those in
(\ref{Idiv:def}), therefore renders a finite result upon integrating
over the entire 3-body phase space. Bose symmetry guarantees that
$I_{{\rm fin}}$ is manifestly symmetric under interchange between
$x_1$ and $x_2=2-z-x_1$. Since its explicit expression is somewhat
lengthy, so will not be reproduced here.

Accordingly, the energy distribution of the parton cross section can
also be decomposed into two parts:
%--------------------------
\bqa
%--------------------------
{d\sigma[e^+e^- \to c\bar{c}({}^1P_1^{(1)},P)+g g]\over dz}=
{d\hat{\sigma}_{\rm div}\over dz} + {d\hat{\sigma}_{\rm fin}\over
dz},
%--------------------------
\eqa
%--------------------------
which are obtained by integrating
(\ref{ampl:squared:gamma*:ccbargg}) over the entire momentum range
of gluon 1:
%--------------------------
\begin{subequations}
%--------------------------
\bqa
%--------------------------
& & \int^{1+r}_{2\sqrt{r}}\! dz \,{d\hat{\sigma}_{\rm div} \over dz}
= {\pi\alpha\over 3 s^2} \int\!d\Phi_3 \,I_{\rm div}(x_1,z),
%--------------------------
\label{dsig:div:dz:def}
%--------------------------
\\
%--------------------------
& & \int^{1+r}_{2\sqrt{r}}\! dz \, {d\hat{\sigma}_{\rm fin} \over
dz} = {\pi\alpha\over 3 s^2} \int\!d\Phi_3 \,I_{\rm fin}(x_1,z).
%--------------------------
\label{dsig:fin:dz:def}
%--------------------------
\eqa
%--------------------------
\end{subequations}
%--------------------------
In deriving these, we have used the conversion formula explained in
footnote~\ref{conversion:decay:production}, as well as included a
factor ${1\over 2!}$ to account for the indistinguishability of two
gluons in the final state.

It is straightforward to complete the integration over $x_1$  in the
right side of (\ref{dsig:fin:dz:def}). Since everything is finite,
this integral can be directly calculated in 4 dimensions.
%----------------------------
In contrast, integrating  $I_{{\rm div}}$ over $x_1$ in DR requires
some special care due to emergence of the IR divergence that occurs
at $z=1+r$. We devote Appendix~\ref{appendix:DR:isolate:IR:div} to
expounding the intermediate technical steps, and here just simply
jump to the desired results:
%--------------------------
\begin{subequations}
%--------------------------
\bqa
%--------------------------
& & {d \hat{\sigma}_{\rm div}\over dz} =
%--------------------------
{128 \pi e_c^2 \alpha^2 C_F \alpha_s^2 \over 3 m_c^2 s^2}
 {c_\epsilon(4\pi)^\epsilon\over \Gamma(1-\epsilon)}
\Bigg\{  (1-r) \delta(1+r-z) \times
%--------------------------
\nn\\
%--------------------------
&& \bigg(- {1\over \epsilon_{\rm
IR}}-2\ln{\frac{\mu^2}{4m_c^2}}+2\ln\frac{(1-\sqrt{r})^2}
{\sqrt{r}}+1\bigg) + \bigg[{1 \over 1+r-z } \bigg]_+ \bigg(
\frac{z^2-4r}{z-2 r}+\sqrt{z^2-4r}\bigg)
%--------------------------
\nn\\
%--------------------------
&&+ {2 z \over z-2 r}- {2-z - \sqrt{z^2-4 r} \over 1+r-z}\Bigg\},
%--------------------------
\label{dsig:div:dz:expression}
%--------------------------
\\
%--------------------------
& & {d \hat{\sigma}_{\rm fin}\over dz}= \frac{256 \pi e_c^2 \alpha^2
C_F \alpha_s^2}{3
   m_c^2 s^2 } {1\over (2-z)^4 (z-2 r)^5}\Bigg\{
  (z-2r) \sqrt{z^2-4 r}\times
\nn\\
%--------------------------
& &  \bigg[ 16 r (-3+2 r-6 r^2-6 r^3 +3 r^4  +6 r^5) +16 r (7+2 r+24
r^2+r^3-20 r^4-2 r^5) z
\nn\\
%--------------------------
   &&
  + 4
   (2-35 r-72 r^2-74 r^3 + 94 r^4 + 25 r^5) z^2-8 (2-21 r-34 r^2+18 r^3+15 r^4) z^3
\nn\\
%--------------------------
   &&
+2 (3-59 r-6 r^2+32 r^3)
   z^4  +(3+23
   r-14 r^2) z^5 -z^6
   \bigg]\nn\\
%--------------------------
   &&
 + \ln  \frac{z-2
   r+\sqrt{z^2-4 r}}{z-2 r-\sqrt{z^2-4
   r}} \, \bigg[
-32 r^2 (3-r +4 r^2+3
   r^3 +2 r^5+5 r^6)\nn\\
%--------------------------
   && +32 r^2 (8+r+14
   r^2+4 r^3+6 r^4+21 r^5+2
   r^6) z \nn\\
%--------------------------
   &&
+8 r (4-37 r-31 r^2-54
   r^3-38 r^4-149 r^5-31 r^6) z^2
\nn\\
%--------------------------
&&
   -8 r (12-25 r-26 r^2-42 r^3-148 r^4 -51
   r^5)
   z^3\nn\\
%--------------------------
   && +2 r (61
-17 r-55 r^2-363
   r^3-186 r^4) z^4-12 r (8+2 r-21
   r^2-17 r^3) z^5\nn\\
%--------------------------
   &&+(1+45 r
-37
   r^2-65 r^3) z^6-(1+5 r-10 r^2) z^7
   \bigg]
   \Bigg\}.
%--------------------------
\label{dsig:fin:dz:expression}
%--------------------------
\eqa
%--------------------------
\label{parton:dsig:dz:expression}
%--------------------------
\end{subequations}
%--------------------------

The appearance of the $\delta$-function and ``+"-function exhibits
some peculiarity of the energy distribution $d \hat{\sigma}_{\rm
div}/dz$ near the maximal $h_c$ energy. As usual, these functions
should be interpreted as the distributions in the mathematical
sense. The $[f(z)]_+$ function in (\ref{dsig:div:dz:expression}) is
defined such that when convoluting it with an arbitrary function
$g(z)$ that is regular at $z=1+r$, one gets
%--------------------------
\bqa
%--------------------------
\int^{1+r}_{2\sqrt{r}}\!dz\,[f(z)]_+ g(z)&=&
\int^{1+r}_{2\sqrt{r}}\!dz\, f(z) (g(z)-g(1+r)).
%--------------------------
\label{plus:function:convolution}
%--------------------------
\eqa
%--------------------------

From (\ref{dsig:div:dz:expression}), we find that the IR singularity
is exactly located at the maximal value of $z$. One certainly
expects that this IR singularity will be swept out once including
the color-octet contribution. Encouragingly, equation
(\ref{color-octet:short-distance:coef:D:dimensions}) implies that
the differential color-octet coefficient $dF^{\rm LH}_8$ is also
proportional to a peaked distribution $\delta(1+r-z)$.

Since $dF^{\rm LH}_8/dz$ is of order $\alpha_s$ only, in order to
match the $O(\alpha_s^2)$ accuracy of the color-singlet parton cross
section, we need incorporate the $O(\alpha_s)$ correction to the
perturbative color-octet NRQCD matrix element. This correction has
already been inferred previously in the related work on quarkonium
production, so we just present the result~\footnote{Note the authors
of Ref.~\cite{Beneke:1998ks} used DR to regularize the UV divergence
but a gluon mass to regulate the IR divergence for the perturbative
color-octet NRQCD matrix element. Here we use DR to regulate both UV
and IR divergences.}:
%--------------------------
\bqa
%--------------------------
\langle{\cal O}^{c\bar{c}}_8(^1S_0)\rangle_{\rm \overline{MS}}=
\langle{\cal
O}^{c\bar{c}}_8(^1S_0)\rangle^{(0)}-\frac{2C_F\alpha_s}{3N_c\pi
m_c^2}\bigg({1 \over \epsilon_{\rm IR}}+\ln 4\pi-\gamma_E\bigg)
\langle{\cal O}^{c\bar{c}}_1(^1P_1)\rangle^{(0)}+\cdots,
%--------------------------
\label{Renormalized:octet:operator:matrix:element}
%--------------------------
\eqa
%--------------------------
where $C_F={N_c^2-1\over 2N_c}$, and $N_c=3$ is the number of the
colors in QCD. Under renormalization, the color-octet operator
${\cal O}^{c\bar{c}}_8(^1S_0)$ mixes with the color-singlet operator
${\cal O}^{c\bar{c}}_1(^1P_1)$ (it also mixes with the color-octet
operator ${\cal O}^{c\bar{c}}_8(^1P_1)$, but whose effect would
arise at higher order in $v$). It is important to note that, the
${\rm \overline{MS}}$-renormalized perturbative matrix element
$\langle {\cal O}^{c\bar{c}}_8(^1S_0)\rangle$ develops a logarithmic
IR divergence.

In passing, it may be worth stressing that, unlike the NRQCD decay
operators, the analogous production operators are no longer the
local 4-fermion operators. In fact, in a series of
work~\cite{Nayak:2005rw,Nayak:2005rt,Nayak:2006fm}, Nayak, Qiu and
Sterman have recently advocated that one must insert the proper
gauge links to the original definitions of the color-octet
production operators~\cite{Bodwin:1994jh} to warrant the gauge
invariance, and the nontrivial effect due to this gauge completion
first shows up at next-to-next-to-leading order in $\alpha_s$ (it
has also been explicitly examined that, at the NLO in $\alpha_s$,
forgoing the gauge completion for the color-octet operators does not
bring in any inconsistency~\cite{Ma:2005cj}). Thus to our purpose,
we are content with staying with the original definition given in
\cite{Bodwin:1994jh}. In this respect, equation
(\ref{Renormalized:octet:operator:matrix:element}) looks very
similar to the analogous formula for the renormalized color-octet
decay operator in NRQCD~\cite{Petrelli:1997ge,Bodwin:2007zf}.

In light of the matching condition in
(\ref{NRQCD:factorization:formula}) for the
$c\bar{c}({}^1P_1^{(1)})$ channel, one can write down the following
equation for $dF^{\rm LH}_1$ to the order $\alpha_s^2$:
%--------------------------
 \bqa
%--------------------------
& &¡¡{d\sigma[e^+e^- \to c\bar{c}({}^1P_1^{(1)})+g g]\over dz} =
{6N_c \over m_c^3} {dF^{\rm LH}_1(\mu)\over dz} {\bigg |}_{\rm
\overline{MS}} -  {128\pi e_c^2\alpha^2 C_F \alpha_s^2 (1-r) \over 3
m_c^2 s^2}\times
%--------------------------
\nn\\
%--------------------------
&& c_\epsilon \bigg[{1 \over \epsilon_{\rm IR}}+
\ln\frac{\pi\mu^2}{m_c^2} -\gamma_E+\ln r
-2\ln(1-r)-\frac{2}{3}\bigg] \delta(1+r-z),
%--------------------------
\label{equation:to:solve:for:dF1}
%--------------------------
\eqa
%--------------------------
where we have explicitly substituted the LO color-octet coefficient
in (\ref{color-octet:short-distance:coef:D:dimensions}) and the NLO
correction to the color-octet matrix element
(\ref{Renormalized:octet:operator:matrix:element}). The subscript
${\rm \overline{MS}}$ reminds that the color-singlet coefficient
$dF^{\rm LH}_1(\mu)$ is determined in accordance with the ${\rm
\overline{MS}}$ factorization scheme.

Substituting the analytic expressions for the differential
$c\bar{c}({}^1P_1^{(1)})$ production cross section, as assembled in
(\ref{parton:dsig:dz:expression}), into
(\ref{equation:to:solve:for:dF1}), one can readily solve the desired
differential color-singlet coefficient:
%--------------------------
\bqa
%--------------------------
& & {dF_1^{\rm LH} (\mu) \over dz} {\bigg |}_{\rm \overline{MS}} =
{64\pi e_c^2\alpha^2 C_F \alpha_s^2 m_c \over 9N_c s^2} \Bigg\{
\bigg(-\ln{\frac{\mu^2}{4m_c^2}}+2\ln\frac{1-\sqrt{r}}
{1+\sqrt{r}}+\frac{1}{3}\bigg) \times
%--------------------------
\nn\\
%--------------------------
&& (1-r) \delta(1+r-z) +\bigg[{1 \over 1+r-z } \bigg]_+\bigg( {z^2
-4 r\over z-2 r}+\sqrt{z^2-4r}\bigg)
%--------------------------
\nn\\
%--------------------------
&&+\frac{2 z}{z-2 r}- {2-z-\sqrt{z^2-4 r} \over 1+r-z} \Bigg\}+
\frac{m_c^3}{6N_c}  {d \hat{\sigma}_{\rm fin} \over dz},
%--------------------------
\label{differential:F1:L.H.}
%--------------------------
\eqa
%--------------------------
where $d \hat{\sigma}_{\rm fin}/dz$ is given in
(\ref{dsig:fin:dz:expression}). Although this short-distance
coefficient is now free of IR singularity, as it should, it now
depends logarithmically on the NRQCD factorization scale $\mu$.
However, when taking into account the $\mu$-dependence of the
color-octet matrix element $\langle \mathcal{O}_8^{h_c}(^1S_0)
\rangle$, the physical $h_c$ production rate in
(\ref{NRQCD:factorization:formula}) in principle does not depend on
this artificial scale.

\subsection{The integrated short-distance coefficients $F^{\rm LH}_1$ and $F^{\rm LH}_8$ }
\label{int:F1:F8:LH}

Inspecting the differential coefficients in
(\ref{color-octet:short-distance:coef:4:dimensions}) and
(\ref{differential:F1:L.H.}), one finds that the distribution of
$h_c$ becomes singular near its maximum energy. However, it was
noticed long ago that NRQCD expansion breaks down near the kinematic
boundary of quarkonium momentum distribution~\cite{Beneke:1997qw},
thus the NRQCD prediction at fixed order is no longer trustworthy.
In order to reliably describe the energy distribution near the
kinematic end point, one should incorporate the effect of
nonperturbative shape function~\cite{Beneke:1999gq}, as well as
resum large Sudakov logarithm~\cite{Fleming:2003gt}, consequently
the quarkonium spectrum near the endpoint will turn over and get
smeared. However, including such refinement is beyond the scope of
this work, which well deserves a dedicated study.

On the other hand, the integrated $h_c$ production rate is much less
sensitive to the abovementioned effects, and NRQCD velocity
expansion is believed to work well for this quantity. Moreover, the
total production rate of the $h_c$ is also experimentally
accessible. Therefore, it is of phenomenological incentive to find
the integrated short-distance coefficients $F^{\rm LH}_1$ and
$F^{\rm LH}_8$.

Integrating (\ref{differential:F1:L.H.}) over $z$ may seem
straightforward, but it turns out to be difficult to obtain the
analytic expression for the part involving $d \hat{\sigma}_{\rm
fin}/dz$. We utilize a simple trick~\cite{Jia:2009np}, i.e.,
restarting from (\ref{dsig:fin:dz:def}), but this time integrating
over $z$ first, then followed by integrating over $x_1$. After some
algebras, we end up with the following integrated short-distance
coefficients:
%--------------------------
\begin{subequations}
%--------------------------
\bqa
%--------------------------
& & F_1^{\rm LH} (\mu)_{\rm \overline{MS}} =  {64\pi e_c^2\alpha^2
C_F \alpha_s^2 m_c \over 9N_c s^2}
(1-r)\bigg[-\ln\frac{\mu^2}{4m_c^2} + 2\ln(1-r) - {65-84r \over
12(1-r)}
%--------------------------
\label{integrated:F1:L.H.}
\nn\\
%--------------------------
&& + {7+7r-9r^2 \over 6(1-r)^2} \ln r +\frac{r (5-7 r) \ln
   ^2 \frac{1+\sqrt{1-r}}{1-\sqrt{1-r}}}{16(1-r)^2}+
   \frac{(14-15r) \ln
    \frac{1+\sqrt{1-r}}{1-\sqrt{1-r}}}{8(1-r)^{3/2}}
   \bigg],
%--------------------------
%--------------------------
\\
%--------------------------
& & F_8^{\rm LH} = {32\pi^2 e_c^2 \alpha^2\alpha_s m_c \over 3 s^2}
(1-r).
%--------------------------
\label{integrated:F8:L.H.}
%--------------------------
\eqa
%--------------------------
\label{integrated:F1:F8:L.H.}
%--------------------------
\end{subequations}
%--------------------------
The color-octet coefficient  $F_8^{\rm LH}$ can be trivially deduced
by integrating (\ref{color-octet:short-distance:coef:4:dimensions})
over $z$.

It is natural to take the factorization scale $\mu$ around $m_c$. We
then find that $F_1^{\rm LH} (\mu)$ becomes negative in most of the
allowed range of $r$ (including $r\approx 0.08$ of phenomenological
interest), except in a narrow window where $r$ is very small. This
has an immediate consequence, that the color-octet channel becomes
indispensable if one wishes to predict a {\it positive} total
production rate for $e^+ e^-\to h_c+$ light hadrons.

It is enlightening to examine the asymptotic behaviors of
(\ref{integrated:F1:F8:L.H.}) in the limit $\sqrt{s}\gg m$:
%--------------------------
\begin{subequations}
%--------------------------
\bqa
%--------------------------
F_1^{\rm LH}(\mu)_{\rm \overline{MS}}\bigg |_{\rm asym} &=& {64\pi
e_c^2\alpha^2 C_F \alpha_s^2 m_c \over 9N_c s^2} \bigg[-{7\over
12}\ln r-\ln\frac{\mu^2}{4m_c^2} - {65 \over 12}+{7\over 2}
\ln{2}\bigg],
%--------------------------
\label{integrated:F1:L.H.:asym}
%--------------------------
\\
%--------------------------
F_8^{\rm LH} \bigg |_{\rm asym} &=& {32\pi^2 e_c^2 \alpha^2\alpha_s
m_c \over 3 s^2}.
%--------------------------
\label{integrated:F8:L.H.:asym}
%--------------------------
\eqa
%--------------------------
\label{integrated:F1:F8:L.H.:asym}
%--------------------------
\end{subequations}
%--------------------------
The total production rate for $e^+e^-\to h_c+$ light hadrons scales
with the center-of-mass energy as $1/s^2$, which is the same as that
for $e^+e^-\to J/\psi+$ light hadrons~\cite{Jia:2009np}.
Nevertheless, it is interesting to note that the leading scaling
violation in (\ref{integrated:F1:L.H.:asym}) is represented by a
single-logarithmic term ($\propto \ln r$), while that in the process
$e^+e^-\to J/\psi+gg$ is by a double-logarithmic term ($\propto
\ln^2 r$)~\cite{Jia:2009np}. Since the newly proposed perturbative
QCD factorization program for quarkonium
production~\cite{Kang:2011mg} is based on $m_c^2/s$ expansion, it is
natural to envisage that refactorizing the higher-twist two-parton
fragmentation function in Ref.~\cite{Kang:2011mg} may provide a
natural framework to identify these logarithms at $O(\alpha_s^2)$
and resum them to all orders in $\alpha_s$.

%--------------------------------------------------------------------
\section{${\bm h}_{\bm c}$ production associated with charmed hadrons}
\label{matching:hc:ccbar}
%--------------------------------------------------------------------

In this section, we investigate the inclusive $h_c$ production
associated with the charmed hadrons at $B$ factories, i.e.,
$e^+e^-\to h_c+X_{c\bar{c}}$. Our central task is again to infer two
short-distance coefficients appearing in the NRQCD factorization
formula (\ref{NRQCD:factorization:formula}). To avoid confusion, we
will always associate a superscript ``Charm" to $F_1$ and $F_8$ in
this section, to distinguish from the analogous coefficients
associated with the process $e^+e^-\to h_c+$ light hadrons in
Sec.~\ref{matching:hc:gg}.

Both color-singlet and octet channels first occur at
$O(\alpha_s^2)$, characterized with the parton processes $e^+e^-\to
c\bar{c}({}^1P_1^{(1)}, {}^1S_0^{(8)})+c\bar{c}$. Since both
channels share the common parton kinematics, we list some useful
3-body phase-space formulas here.

We denote the momenta of the $c\bar{c}$ pair, $c$, and $\bar{c}$ by
$P$, $k_1$, $k_2$, respectively, with $P^2\approx 4m_c^2$,
$k_1^2=k_2^2=m_c^2$. Analogous to
(\ref{three:energy:fractions:def}), we also introduce three
fractional energy variables $z$, $x_1$ and $x_2$, which are subject
to the constraint $x_1+x_2+z=2$.

Since no massless particles are involved in the final state, the
parton cross sections in both channels do not exhibit any IR
singularity. Unlike in Sec.~\ref{matching:hc:gg}, we thus can
perform the calculation directly in 4 dimensions. Consequently,
suffice it to know the 4-dimensional 3-body phase space measure for
the process $\gamma^*\to c\bar{c}(P)+c(k_1)+\bar{c}(k_2)$:
%--------------------------
\bqa
%--------------------------
\int\! d\Phi_3 &=& {s\over 128\pi^3} \int^{1}_{2\sqrt{r}} d z
\int^{x_1^+}_{x_1^-} dx_1.
%--------------------------
%--------------------------
\eqa
%--------------------------
The integration limits of $z$ have been explicitly specified, while
those for the fractional energy of $c$, $x_1^\pm$, read
%--------------------------
\bqa
%--------------------------
x_1^\pm &=& {2-z\over 2} \pm {1\over 2} \sqrt{(1-z)(z^2-4r)\over
1+r-z}.
%--------------------------
\eqa
%--------------------------

\subsection{Deducing $d F_1^{\rm Charm}$}

We start with calculating the differential color-singlet coefficient
$d F_1^{\rm Charm}$. At the lowest order in $\alpha_s$, only four
Feynman diagrams need to be considered for the process $\gamma^* \to
c\bar{c}({}^1P_1^{(1)})+c\bar{c}$. This calculation is quite similar
to the analogous one for $e^+e^-\to
J/\psi+c\bar{c}$~\cite{Cho:1996cg} and $e^+e^-\to
\chi_{cJ}(\eta_c)+c\bar{c}$ ($ J=0,1,2$)~\cite{Liu:2003jj}. The
perturbative matching calculation for this coefficient is rather
straightforward, so we directly present the result:
%--------------------------
\bqa
%--------------------------
&& {d F_1^{\rm Charm}\over dz}=  {64 \pi e_c^2\alpha^2
 \alpha_s^2
  \over 243 m_c s (2-z)^4 z^4}
   \Bigg\{
   \frac{\sqrt{(1+r-z)(1-z) (z^2-4 r)}}{(2-z)^4}
    \bigg[768 r^4 -384 r^3 (8+5 r) z \nn\\
%--------------------------
&&   - 64
   r (8-16 r-128 r^2-35 r^3) z^2+32
   r (56-64 r-310 r^2-43 r^3) z^3 \nn\\
%--------------------------
    && +16 (8-140 r +136 r^2+438 r^3+25
   r^4) z^4-8 (48-80 r+136 r^2+336
   r^3-15 r^4) z^5\nn\\
%--------------------------
    &&+4 (152+164 r-112
   r^2+132 r^3-7 r^4) z^6-(672+560
   r-400 r^2+84 r^3-6 r^4) z^7
   \nn\\
%--------------------------
    &&+2 (300+182 r-60 r^2-r^3) z^8-8 (49+14 r-2
   r^2) z^9+(130+17 r) z^{10}-18 z^{11}\bigg]\nn\\
%--------------------------
    &&
  -\frac{r}{2 z}   \ln \frac{z\sqrt{1+r-z}+\sqrt{(1-z) (z^2-4
   r)}}{z\sqrt{1+r-z}-\sqrt{(1-z) (z^2-4
   r)}} \times \nn\\
%--------------------------
    &&\bigg[ -192 r^4+96 r^3 (8+3 r) z+32 r (4-8
   r-39 r^2-6 r^3) z^2-32 r (10-4
   r-26 r^2-r^3) z^3\nn\\
%--------------------------
    && -4 (8-132
   r-44 r^2+48 r^3+5 r^4) z^4 + 2 (48-192
   r-64 r^2+48 r^3+3 r^4) z^5\nn\\
%--------------------------
    &&  -2 (72+36
   r+82 r^2+11 r^3) z^6 +4 (38+61
   r +13 r^2) z^7-(90+59 r) z^8+16 z^9
  \bigg]
   \Bigg\}.
%--------------------------
\label{differential:F1:charm}
%--------------------------
\eqa
%--------------------------

\subsection{Deducing $d F_8^{\rm Charm}$}

Next we proceed to calculate the differential color-octet
coefficient $d F_8^{\rm Charm}$. At the order $\alpha_s^2$, we need
consider six Feynman diagrams for the parton process $\gamma^*\to
c\bar{c}({}^1S_0^{(8)})+c\bar{c}$. The perturbative matching for
this coefficient is analogous to that in Sec.~\ref{match:hc:gg:dF8},
which is also quite straightforward. Here we just give the result:
%--------------------------
\bqa
%--------------------------
& & {d F_8^{\rm Charm} \over dz} =  {\pi e_c^2\alpha^2 \alpha_s^2
\over 27 m_c s } {1\over (1+r-z)(2-z)^2 z^3}\Bigg\{\frac{2 z
\sqrt{(1-z) (z^2-4 r)}}{3 (2-z)^4
   (1+r-z)^{3/2}} \times
%--------------------------
\nn\\
%--------------------------
& & \bigg[ 96 r^3 (1+r)^3
   -96 r^2 (1+r)^2 (4+8
   r+r^2) z\nn\\
%--------------------------
   &&+16 (6+59
   r
+186 r^2+328 r^3+284 r^4+51
   r^5-2 r^6) z^2
   \nn\\
%--------------------------
& & -8 (48+374 r+770
   r^2+965 r^3+551 r^4+41 r^5-r^6 ) z^3\nn\\
%--------------------------
   &&
 +(592+3392 r+5556 r^2+5806 r^3+2234
   r^4+90 r^5 -6 r^6) z^4 \nn\\
%--------------------------
   && - 2
   (232+704
   r+1162 r^2+1131 r^3+244 r^4-5 r^5) z^5
\nn\\
%--------------------------
   && +(262-207 r+217 r^2 +293 r^3+11
   r^4 ) z^6
   -(184-397 r-186 r^2-15 r^3) z^7
    \nn\\
%--------------------------
   && +(116-139 r-40 r^2)z^8
  -(40-13 r) z^9
  + 6
   z^{10}\bigg]  + r \ln \frac{z\sqrt{1+r-z}+\sqrt{(1-z) (z^2-4
   r)}}{z\sqrt{1+r-z}-\sqrt{(1-z) (z^2-4
   r)}}\times
%--------------------------
\nn\\
%--------------------------
 && \bigg[
    8 r^3 (1+r)-8
   r^2 (4+5 r) z-2 (4-2 r-22 r^2-r^3-r^4) z^2
  \nn\\
  %--------------------------
   && +(8+8 r+40 r^2-6 r^3)
   z^3 -(14+33
   r-5 r^2) z^4
+(2-19 r) z^5+
  12 z^6
   \bigg]
   \Bigg\}.
%--------------------------
\label{differential:F8:charm}
%--------------------------
\eqa
%--------------------------

\subsection{Fragmentation function for $c\to h_c$ and integrated cross section}
\label{Frag:Func:c:hc:int:X:section}

At first sight, the differential coefficients in
(\ref{differential:F1:charm}) and (\ref{differential:F8:charm}) may
look too disordered to extract anything useful. However, this is
just a disguise, since we are certain that in the asymptotic limit
$\sqrt{s}\gg m_c$, the differential $h_c$ production rate associated
with $c\bar{c}$ must be dominated by the fragmentation mechanism:
%--------------------------
\bqa
%--------------------------
{d\sigma[e^+e^-\to h_c(P)+ X_{c\bar{c}}]\over dz}  &=& 2
\check{\sigma} D_{c\to h_c}(z),
%--------------------------
\label{frag:func:def:e+e-:annih}
%--------------------------
\eqa
%--------------------------
where $\check{\sigma}= N_c {4\pi e_c^2\alpha^2 \over 3 s}$ is the
cross section for the process $e^+e^-\to c\bar{c}$, and $D_{c\to
h_c}(z)$ stands for the fragmentation probability for $c$ into the
$h_c$ carrying the energy fraction $z$. The factor 2 arises because
both $c$ and $\bar{c}$ can fragment into $h_c$ with equal
probability.

According to the NRQCD factorization ansatz, the fragmentation
function of $c$ into $h_c$ can be refactorized as
%--------------------------
\bqa
%--------------------------
D_{c\to h_c}(z) =  d_1^{h_c}(z) {\langle
\mathcal{O}_1^{h_c}(^1P_1)\rangle \over m_c^5} + d_8^{h_c}(z)
{\langle \mathcal{O}_8^{h_c}(^1S_0) \rangle \over m_c^3},
%--------------------------
\label{Fragmentation:function:NRQCD:factorization}
%--------------------------
\eqa
%--------------------------
where $d_1^{h_c}(z)$ and $d_8^{h_c}(z)$ are the corresponding
short-distance coefficient (jet) functions.

Comparing (\ref{frag:func:def:e+e-:annih}) with
({\ref{NRQCD:factorization:formula}}), one can immediately realize
that, these short-distance functions can be obtained by taking the
asymptotic ($r\to 0$) limit of $d F^{\rm Charm}_n/dz$ ($n=1,8$):
%--------------------------
\begin{subequations}
%--------------------------
\bqa
%--------------------------
 d_1^{h_c}(z)&=&{m_c\over 2 \check{\sigma}} {d F_1^{\rm Charm}
\over dz}\bigg|_{\rm asym}
%--------------------------
\nn \\
%--------------------------
&=&  {16 \alpha_s^2 z (1-z)^2 (64-128 z+176 z^2-160 z^3+140 z^4-56
z^5+9 z^6) \over 243(2-z)^8},
%--------------------------
\label{d1:short-dist:coef:frag:func}
%--------------------------
\\
%--------------------------
d_8^{h_c}(z)&=& {m_c\over 2 \check{\sigma}} {d F_8^{\rm Charm} \over
dz}\bigg|_{\rm asym} = {\alpha_s^2 z (1-z)^2 \left(48+8 z^2-8 z^3+ 3
z^4 \right)\over 162(2-z)^6}.
%--------------------------
\label{d8:short-dist:coef:frag:func}
%--------------------------
\eqa
%--------------------------
\label{d1:d8:short-dist:coef:frag:function}
%--------------------------
\end{subequations}
%--------------------------
Both the expressions for $d_1^{h_c}(z)$ and $d_8^{h_c}(z)$ in
(\ref{d1:d8:short-dist:coef:frag:function}) fully agree with
Ref.~\cite{Yuan:1994hn}.

Honestly speaking, the $B$ factory energy is far from being
asymptotically large. Therefore for this phenomenologically relevant
case, the fragmentation function calculated in
(\ref{d1:d8:short-dist:coef:frag:function}) can hardly faithfully
reproduce the energy distribution of the $h_c$ depicted in
(\ref{differential:F1:charm}) and (\ref{differential:F8:charm}).

Like what has been done in Sec.~\ref{int:F1:F8:LH}, it is also of
both theoretical and phenomenological interest in knowing the
integrated production rate for $e^+e^-\to h_c+$ charmed hadrons.
Conceivably, it seems extremely challenging, if not impossible, to
complete the integration of (\ref{differential:F1:charm}) and
(\ref{differential:F8:charm}) over $z$ in closed form. Nevertheless,
it is quite easy to infer the asymptotic behavior of the total cross
section with the help of the fragmentation function:
%--------------------------
\bqa
%--------------------------
& & \sigma[e^+e^-\to h_c+ X_{c\bar{c}}]\bigg|_{\rm asym} =
2\check{\sigma} \int^1_0\!dz\,D_{c\to h_c}(z)
%--------------------------
\nn\\
%--------------------------
& & = 16 \alpha_s^2 \check{\sigma}  \bigg[  {  18107- 26110\ln 2
\over 8505} {\langle \mathcal{O}_1^{h_c}(^1P_1)\rangle \over m_c^5}+
{  773-1110 \ln 2 \over 12960} {\langle \mathcal{O}_8^{h_c}(^1S_0)
\rangle \over m_c^3} \bigg].
%--------------------------
\label{hc:c:cbar:tot:cross:section:asym}
%--------------------------
\eqa
%--------------------------
In contrast to (\ref{integrated:F1:F8:L.H.:asym}), the production
rate for $h_c+$ charmed hadron exhibits much slower asymptotic
decrease ($\propto 1/s$) than that for $h_c+$ light hadrons
($\propto 1/s^2$), which clearly corroborates the dominance of the
fragmentation mechanism at high energy ($p_T$).

%--------------------------------------------------------------------
\section{Phenomenology}
\label{phenomenology}
%--------------------------------------------------------------------

With various color-singlet and color-octet short-distance
coefficients determined in Secs.~\ref{matching:hc:gg} and
\ref{matching:hc:ccbar}, we are ready to make a concrete analysis
for the inclusive $h_c$ production at $B$ factories and assess its
observation prospects.

\subsection{Input parameters}

At the $B$ factory energy, we choose the running QED coupling
constant $\alpha(\sqrt{s})=1/131$, and the running QCD coupling
constant $\alpha_s(\sqrt{s}/2)\approx 0.211$. For the process
$e^+e^- \to h_c+$ light hadrons, we take the occurring factorization
scale $\mu=m_c$. An important source of the uncertainty in our
predictions is rooted in the uncertainty about charm quark mass. We
take the customary value $m_c=1.5$ GeV, but allow it to float
between 1.3 to 1.8 GeV in order to closely assess this uncertainty.

We need also specify the values of two nonperturbative NRQCD matrix
elements appearing in the factorization formula
(\ref{NRQCD:factorization:formula}). Upon vacuum saturation
approximation, one can relate the color-singlet matrix element with
the square of the first derivative of the radial wave function at
the origin for $1P$ charmonium, which is calculable in quark
potential models~\cite{Bodwin:1994jh}:
%--------------------------
\bqa
%--------------------------
\langle\mathcal{O}_1^{h_c}(^1P_1)\rangle \approx \frac{3}{2J+1}
\langle\mathcal{O}_1^{\chi_{cJ}}(^3P_J)\rangle \approx {9N_c \over
2\pi}|R_{1P}^\prime(0)|^2.
%--------------------------
\eqa
%--------------------------
If the value of $R_{1P}^\prime(0)$ is calculated from the
Buchm\"{u}ller-Tye potential model~\cite{Eichten:1995ch}, one then
finds $\langle \mathcal{O}_1^{h_c}(^1P_1)\rangle=0.322\;{\rm
GeV}^{5}$.

In contrast to the color-singlet matrix element, no reliable
phenomenological model calculations are available for the
color-octet matrix element
$\langle\mathcal{O}_8^{h_c}(^1S_0)\rangle$, since neither vacuum
saturation approximation nor the quark potential models are
applicable in this situation.

Fortunately, the approximate heavy quark spin symmetry in NRQCD can
be invoked to connect the color-octet matrix elements of $h_c$ and
$\chi_{cJ}$ ($J=0,1,2$)~\cite{Bodwin:1994jh}:
%--------------------------
 \bqa
 %--------------------------
\langle\mathcal{O}_8^{h_c}(^1S_0)\rangle\approx
\frac{3}{2J+1}\langle\mathcal{O}_8^{\chi_{cJ}}(^3S_1)\rangle.
%--------------------------
\label{CO:ME:heavy:quark:spin:symm:hc:chicJ}
%--------------------------
\eqa
%--------------------------
There have been available a number of phenomenological studies for
inclusive $\chi_{cJ}$ production in hadron collision or in $B$ decay
experiments~\cite{Cho:1995ce,Braaten:1994vv,Beneke:1998ks,Ma:2010vd,Gong:2012ug,Cacciari:1995yt},
and the color-octet matrix elements
$\langle\mathcal{O}_8^{\chi_{cJ}}(^3S_1)\rangle$ have been fitted by
various groups over years. We can use
(\ref{CO:ME:heavy:quark:spin:symm:hc:chicJ}) to translate their
fitted color-octet matrix elements for $\chi_{cJ}$ to the desired
one for $h_c$. In Table~\ref{Table:1}, we have enumerated some
values of $\langle\mathcal{O}_8^{h_c}(^1S_0)\rangle$ excerpted from
various references.

%-------------------------------------------------------------------
\begin{table}[tbp]
%\begin{table}[!hbp]
\centering \caption{Variation of the predicted total cross sections
for $e^+e^-\to h_c+$ light hadrons and for $e^+e^-\to h_c+$ charmed
hadrons with the value of $\langle\mathcal{O}_8^{h_c}(^1S_0)
\rangle$ (in units of ${\rm GeV}^3$). We have taken $\alpha=1/131$,
$\alpha_s(\sqrt{s}/2)=0.211$, $m_c=1.5$ GeV, and the color-singlet
matrix element $\langle {\cal O}_1(^1P_1)\rangle=0.322\;{\rm
GeV}^5$. The color-octet matrix element is taken from various
references, while the last entry provides a lower bound inferred
from the renormalization-group equation running in
(\ref{CO:ME:estimate:RGE:running}). \label{Table:1} }
%---------------------------------------------------
\begin{tabular}{|c|c|c|c|c|c|}
%---------------------------------------------------
\hline \multicolumn{1}{|c|}{\multirow{2}*{\backslashbox{\;\;\;\;
Cross Section} {\!\!\!\!\!$\langle \mathcal{O}_8^{h_c}(^1S_0)
\rangle$ }}}& Ref.~\cite{Cho:1995ce,Braaten:1994vv}&
Ref.~\cite{Cacciari:1995yt} & Ref.~\cite{Beneke:1998ks,Gong:2012ug}
& Ref.~\cite{Ma:2010vd} & RGE
%---------------------------------------------------
\\
%---------------------------------------------------
\cline{2-6} &$0.009-0.01$&$0.022-0.025$ &$0.014-0.020$&0.039&$\ge
0.0085$
%---------------------------------------------------
\\
%---------------------------------------------------
\hline $e^+e^-\to h_c+X_{\rm LH}$ (fb)& $86.7-97.6$ & $229.2-262.0$
& $141.5-207.2$ & 415.5& $\ge 81.2$
%---------------------------------------------------
\\
%---------------------------------------------------
\hline $e^+e^-\to h_c+X_{c\bar{c}} $ (fb) &
$9.9-10.0$&$10.2-10.3$&$10.1-10.2$&10.7&$\ge 9.9$
%---------------------------------------------------
\\
%---------------------------------------------------
\hline
\end{tabular}
\end{table}
%-------------------------------------------------------------------

Alternatively, one can infer an order-of-magnitude estimate for the
lower bound on $\langle\mathcal{O}_8^{h_c}(^1S_0)\rangle$ using the
renormalization group equation (RGE), which governs the
renormalization scale dependence of this color-octet matrix element.
The solution to the RGE at leading order in $\alpha_s$
reads~\cite{Bodwin:1994jh}
%--------------------------
\bqa
%--------------------------
\langle {\cal O}^{h_c}_8({}^1S_0)\rangle_{m_c}  = \langle {\cal
O}^{h_c}_8(^1S_0)\rangle_\mu  + {8 C_F \over 3N_c\beta_0}\ln
\bigg(\frac{\alpha_s(\mu)} {\alpha_s(m_c)}\bigg)\frac{\langle {\cal
O}^{h_c}_1(^1P_1)\rangle}{m_c^2},
%--------------------------
\eqa
%--------------------------
where $\beta_0= { 11N_c-2 n_f \over 3}=9$ is the one-loop
coefficient of QCD $\beta$ function with $n_f=3$ light quark
flavors. The subscript of the color-octet matrix element specifies
the renormalization scale affiliated with the operator ${\cal
O}^{h_c}_8(^1S_0)$. Taking $\mu=m_c v$, and assuming the matrix
element $\langle {\cal O}^{h_c}_8(^1S_0)\rangle_{m_c v}$ is
nonnegative, one then gets:
%--------------------------
\bqa
%--------------------------
\langle {\cal O}_8(^1S_0)\rangle_{m_c}
\ge\frac{32}{243}\ln\bigg(\frac{\alpha_s(m_cv)}{\alpha_s(m_c)}\bigg)
{\langle {\cal O}_1(^1P_1)\rangle \over m_c^2}.
%--------------------------
\label{CO:ME:estimate:RGE:running}
%--------------------------
\eqa
%--------------------------
Taking $\alpha_s(m_c)\sim 0.35$ and $\alpha_s(m_c v)\sim v\sim
0.55$, we then obtain $\langle {\cal O}_8(^1S_0)\rangle_{m_c}
\gtrsim 0.0085\;{\rm GeV}^3$. This estimate is of course a very
rough one. Nevertheless, as one can clearly see from
Table~\ref{Table:1}, all the phenomenologically determined values
for the matrix element $\langle {\cal
O}^{h_c}_8(^1S_0)\rangle_{m_c}$ seem to be compatible with this
bound.

In the following numerical analyses, we will take $\langle {\cal
O}^{h_c}_8(^1S_0)\rangle_{m_c}=0.02\;{\rm GeV}^3$, a medium value
among those tabulated in Table~\ref{Table:1}.

\subsection{Numerical results}

%--------------------------
\begin{figure}
\begin{center}
\includegraphics[height=5 cm]{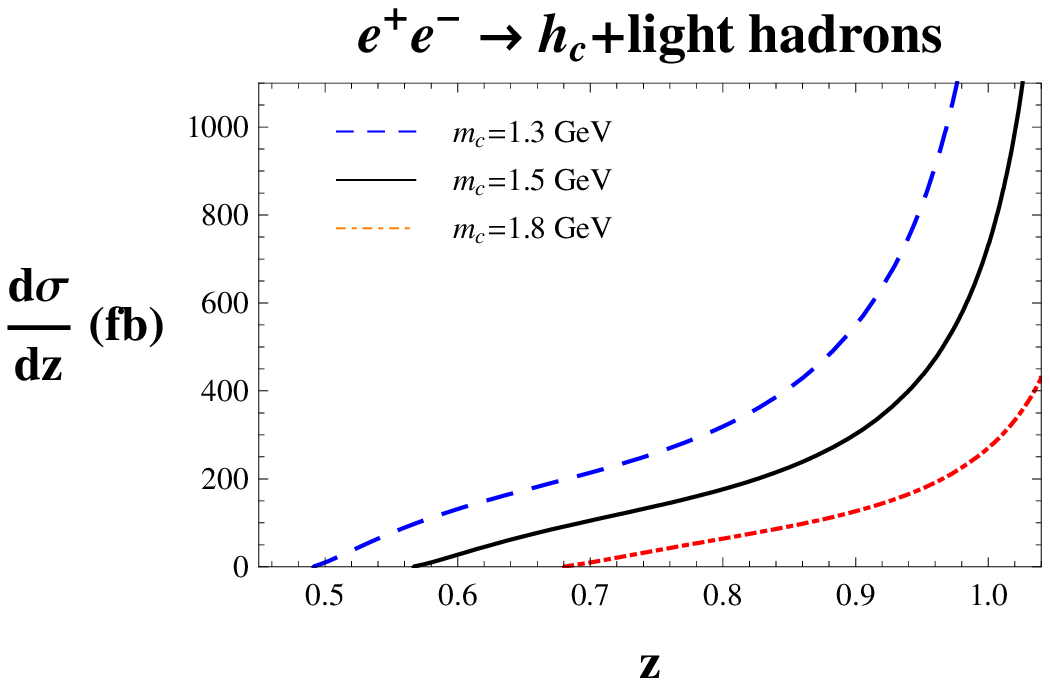}
\includegraphics[height=5 cm]{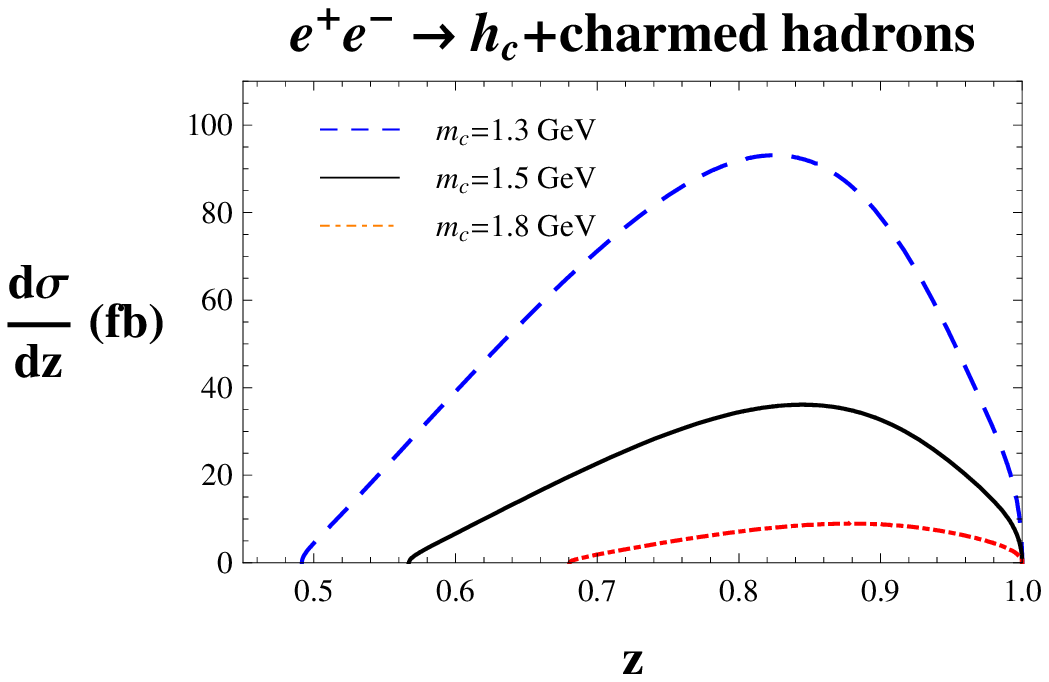}
\caption{The energy distribution of $h_c$ in the processes
$e^+e^-\to h_c+X_{\rm LH}$ (left panel) and $e^+e^-\to h_c+
X_{c\bar{c}}$ (right panel) at $\sqrt{s}=10.58$ GeV. The color-octet
matrix element $\langle\mathcal{O}_8^{h_c}(^1S_0)\rangle$ is taken
as $0.02\;{\rm GeV}^3$. Three curves in each plot correspond to
taking $m_c=$ 1.3, 1.5, and 1.8 GeV, respectively.
\label{hc:energy:distribution:plot}}
\end{center}
\end{figure}
%--------------------------

Substituting the differential short-distance coefficients $dF_n^{\rm
LH}$ $(n=1,8)$ given in
(\ref{color-octet:short-distance:coef:4:dimensions}) and
(\ref{differential:F1:L.H.}) into
(\ref{NRQCD:factorization:formula}), we obtain the energy
distribution of $h_c$ in the process $e^+e^-\to h_c+$ light hadrons
at $B$ factory energy, as shown in the left panel of
Fig.~\ref{hc:energy:distribution:plot}. Similarly, substituting the
differential coefficients $dF_n^{\rm Charm}$ $(n=1,8)$ given in
(\ref{differential:F1:charm}) and (\ref{differential:F8:charm}) into
(\ref{NRQCD:factorization:formula}), we then obtain the energy
spectrum of $h_c$ in the process $e^+e^-\to h_c+$ charmed hadrons,
as shown in the right panel of
Fig.~\ref{hc:energy:distribution:plot}.

As can be clearly seen from Fig.~\ref{hc:energy:distribution:plot},
the $h_c$ energy spectra in two production channels are markedly
different. In the former case, the differential production rate of
$h_c$ sharply rises and diverges at the maximal energy of $h_c$;
while in the latter, the distribution turns over and vanishes as the
energy of $h_c$ approaches its maximum.  As was discussed in
Sec.~\ref{int:F1:F8:LH}, for the $h_c$ production associated with
light hadrons, our prediction to the high-$z$ part of the $h_c$
spectrum becomes untrustworthy, due to the breakdown of perturbative
and velocity expansions near the endpoint region. An appropriate
treatment requires resumming Sudakov logarithms as well as
incorporating nonperturbative shape function, consequently one then
expects the $h_c$ spectrum will be smeared, and turn over near the
upper endpoint, rather than diverge.

As we will see shortly, this process is largely dominated by the
color-octet channel rather than the -singlet channel. From
(\ref{color-octet:short-distance:coef:4:dimensions}), the $h_c$
spectrum in color-octet channel at LO in $\alpha_s$ is a sharp
$\delta$-function spiked on the maximal $h_c$ energy. Even though
taking into account that the radiation of soft gluons would smear
the $h_c$ energy spectrum in this channel, it is reasonable to
expect that the majority of the $h_c$ events will still be located
near the upper end of the momentum spectrum. This may serve as some
useful guidance for the experimentalists.

Our predictions for the integrated cross sections should be much
more reliable than the differential distributions. Substituting the
integrated short-distance coefficients $F_n^{\rm LH}$ $(n=1,8)$,
which are collected in (\ref{integrated:F1:F8:L.H.}), into
(\ref{NRQCD:factorization:formula}), we obtain the total $h_c$
production rate for the process $e^+e^-\to h_c+$ light hadrons; for
the channel $e^+e^-\to h_c+$ charmed hadrons, the total cross
section can be reached by numerically integrating the respective
differential distribution.

With our default value for the color-octet matrix element
$\langle\mathcal{O}_8^{h_c}(^1S_0)\rangle = 0.02\;{\rm GeV}^3$, we
estimate the total cross section for $h_c+$ light hadrons to be
about 207 fb, and that for $h_c+$ charmed hadrons to be about 10 fb.
In Table~\ref{Table:1}, we tabulate various predictions for the
$h_c$ total cross sections in both production channels by adopting
different values of color-octet matrix elements, which are in the
following ranges:
%--------------------------
\begin{subequations}
%--------------------------
\bqa
%--------------------------
&&\sigma[e^+e^-\to h_c +X_{\rm LH}] = 81.2-415.5 \;{\rm fb},
%--------------------------
\label{predictions:total:hc:crossX:chan:LH}
%--------------------------
\\
%--------------------------
&&\sigma[e^+e^-\to h_c+X_{c\bar{c}}] = 9.9-10.7 \;{\rm fb}.
%--------------------------
\eqa
%--------------------------
\label{predictions:total:hc:cross:sec:two:chan}
%--------------------------
\end{subequations}
%--------------------------
One immediately observe that, the cross section of the former
channel is quite sensitive to the value of
$\langle\mathcal{O}_8^{h_c}(^1S_0)\rangle$, but that of the latter
is not sensitive to it at all. This clearly indicates that the
former process is dominated by the color-octet channel, while the
latter is dominated by the color-singlet channel.

It is interesting to contrast the inclusive $h_c$ production at $B$
factories with the analogous inclusive $J/\psi$ production
processes, which have been recently measured by \textsc{Belle}
Collaboration~\cite{Pakhlov:2009nj}:
%--------------------------
\begin{subequations}
%--------------------------
\bqa
%--------------------------
&&\sigma[e^+e^-\to J/\psi+X_{\rm LH}] = 430 \pm 90 \pm 90 \;{\rm
fb},
%--------------------------
\label{J/psi:light:hadrons:obs:cross:sec}
%--------------------------
\\
%--------------------------
&&\sigma[e^+e^-\to J/\psi+X_{c\bar{c}}] = 740 \pm 80^{+90}_{-80}
\;{\rm fb}.
%--------------------------
\eqa
%--------------------------
\end{subequations}
%--------------------------
We see that the cross section for $h_c+$ light hadrons is comparable
in magnitude with the observed production rate for $J/\psi+$ light
hadrons. In this work we have only implemented the color-octet
short-distance coefficient $F_8^{\rm LH}$ at LO in $\alpha_s$. It
was recently discovered that there may exist a large positive
$O(\alpha_s)$ correction to this coefficient~\cite{Zhang:2009ym}. If
we include this perturbative correction, the total production rate
for $h_c+$ light hadrons may easily reach the value given in
(\ref{J/psi:light:hadrons:obs:cross:sec}). Therefore, copious $h_c+$
light hadrons events should already have been produced at $B$
factories.

One naturally expects that the NRQCD factorization framework will
eventually break down as one keeps pushing down the center-of-mass
energy. Leaving this caveat aside, it may still be tempting to
bluntly apply our formula to the process $e^+e^-\to h_c \pi^+\pi^-$
at $\sqrt{s}=4.17$ GeV, first observed in the \textsc{CLEO}
experiment~\cite{CLEO:2011aa}. Taking the default value of the
color-octet matrix element $\langle\mathcal{O}_8^{h_c}(^1S_0)\rangle
= 0.02\;{\rm GeV}^3$, and choosing $\alpha_s(\sqrt{s}/2)=0.305$,
$m_c=1.5$ GeV, we then obtain $\sigma[e^+e^-\to h_c+X_{\rm
LH}]\approx 2.5$ pb, which is considerably lower than the measured
cross section $15.6\pm2.3\pm1.9\pm 3.0$ pb~\cite{CLEO:2011aa}. This
may indicate that, when $\sqrt{s}$ gets close to the open charm
threshold, $\sqrt{s}\simeq 2 m_c$, the $h_c$ production is almost
saturated by the exclusive events, therefore the NRQCD
factorization, which is tailor-made to tackle the inclusive
quarkonium production, becomes inevitably untrustworthy. To handle
this situation more appropriately, one likely needs appeal to the
even lower-energy effective field theory such as the potential
NRQCD.

In sharp contrast with inclusive $J/\psi$ production, our predicted
production rate for $h_c+$ charmed hadrons at the $B$ factories is
about one order of magnitude smaller than that for $h_c+$ light
hadrons. This is a normal hierarchy pattern, being consistent with
the heuristic expectation based on the kinematics consideration.
Practically speaking, the low production rate may render the
experimental measurements of this production channel of $h_c$
difficult.

%--------------------------
\begin{figure}
\begin{center}
\includegraphics[height= 5 cm]{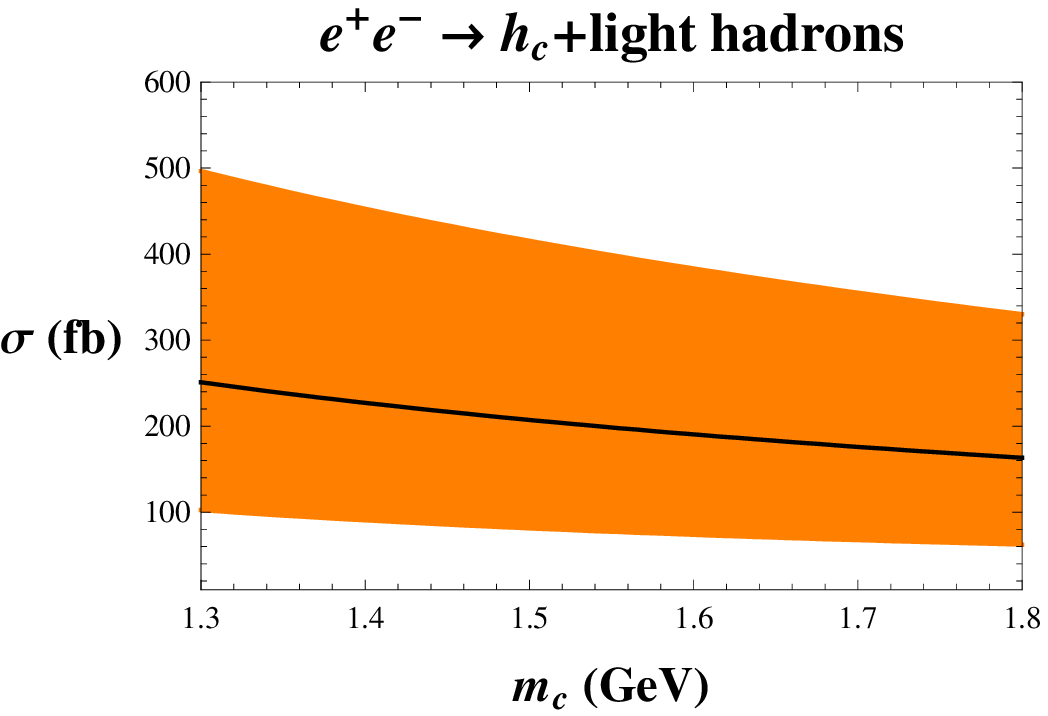}
\includegraphics[height= 5 cm]{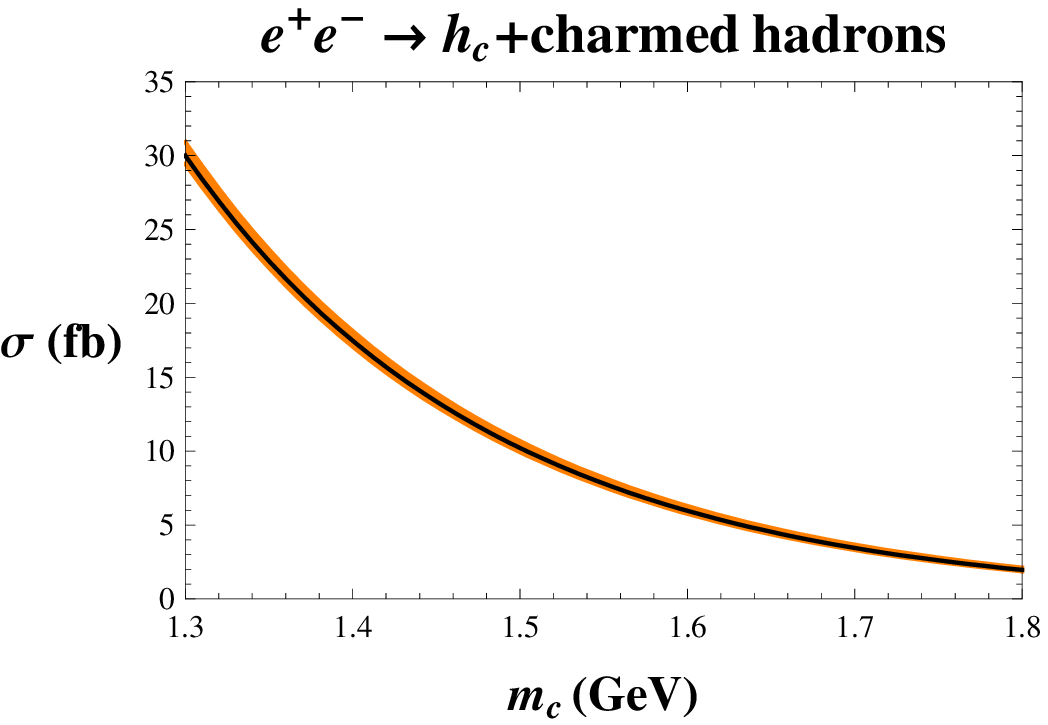}
\caption{The total production rate for $h_c$ as a function of $m_c$
at $\sqrt{s}=10.58$ GeV. The left panel is for the process
$e^+e^-\to h_c+X_{\rm LH}$, and the right panel is for $e^+e^-\to
h_c+ X_{c\bar{c}}$. The band in each plot characterizes the
uncertainty estimated by varying the color-octet matrix element
$\langle\mathcal{O}_8^{h_c}(^1S_0)\rangle$ from 0.0085 to 0.039
${\rm GeV}^3$, where the central curve corresponds to fixing
$\langle\mathcal{O}_8^{h_c}(^1S_0)\rangle$ at $0.02\;{\rm GeV}^3$.
\label{hc:against:mc:plot}}
\end{center}
\end{figure}
%--------------------------

One should caution that, the predicted $h_c$ production rates given
in (\ref{predictions:total:hc:cross:sec:two:chan}) are still subject
to large theoretical uncertainties. Besides the value of
$\langle\mathcal{O}_8^{h_c}(^1S_0)\rangle$, the charm quark mass
constitutes one major source of the uncertainties.  In
Fig.~\ref{hc:against:mc:plot} we explicitly show how the total
production rates for $h_c$ vary with $m_c$. A useful message
conveyed by Fig.~\ref{hc:against:mc:plot} is that, the production
rate for $e^+e^-\to h_c+$ charmed hadrons is much more sensitive to
$m_c$ than that for $e^+e^-\to h_c+$ light hadrons (this point can
also be seen in Fig.~\ref{hc:energy:distribution:plot}). This fact
can be best explained by going to the asymptotic limit $\sqrt{s}\gg
m_c$. In such a limit, the process $e^+e^-\to h_c+X_{c\bar{c}}$ is
dominated by the color-singlet fragmentation mechanism, thus in
light of (\ref{hc:c:cbar:tot:cross:section:asym}), one expects
$\sigma\propto \langle O^{h_c}_1({}^1P_1)\rangle /m_c^5$.  The
process $e^+e^-\to h_c+X_{\rm LH}$ is dominated by the color-octet
channel. From (\ref{integrated:F8:L.H.:asym}), one finds that,
asymptotically $\sigma \propto \langle O_8(^1S_1)\rangle/m_c$. It is
this very different power-law scaling behavior that accounts for the
much stronger sensitivity of $\sigma[e^+e^-\to h_c+X_{c\bar{c}}]$ to
charm quark mass.

%--------------------------
\begin{figure}
\begin{center}
\includegraphics[height= 5 cm]{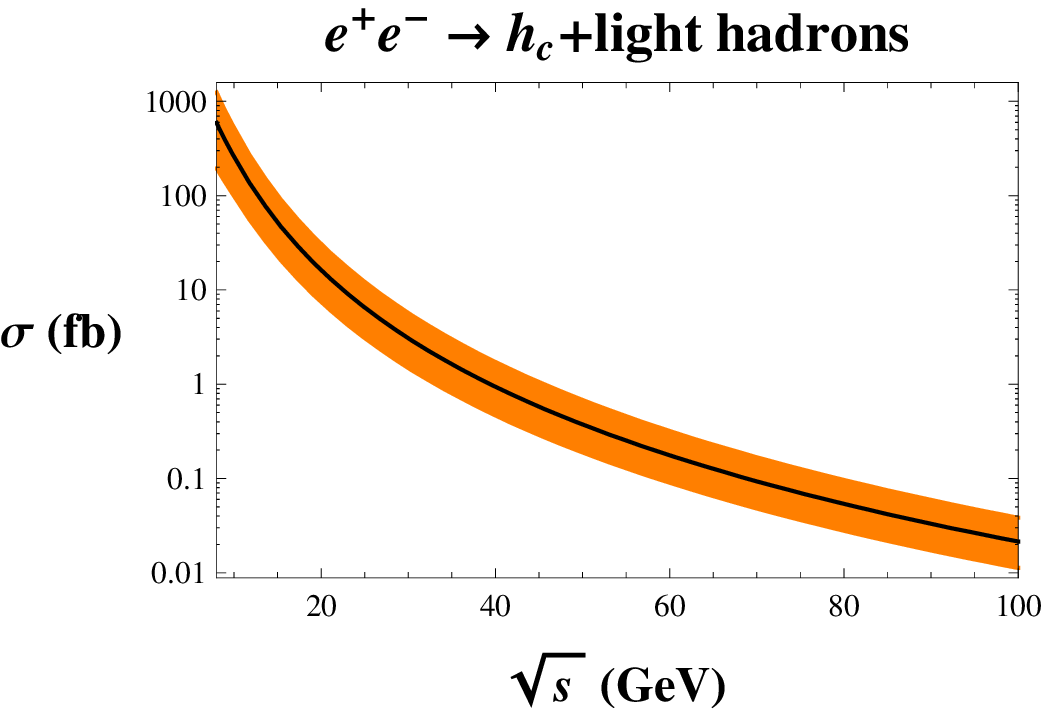}
\includegraphics[height= 5 cm]{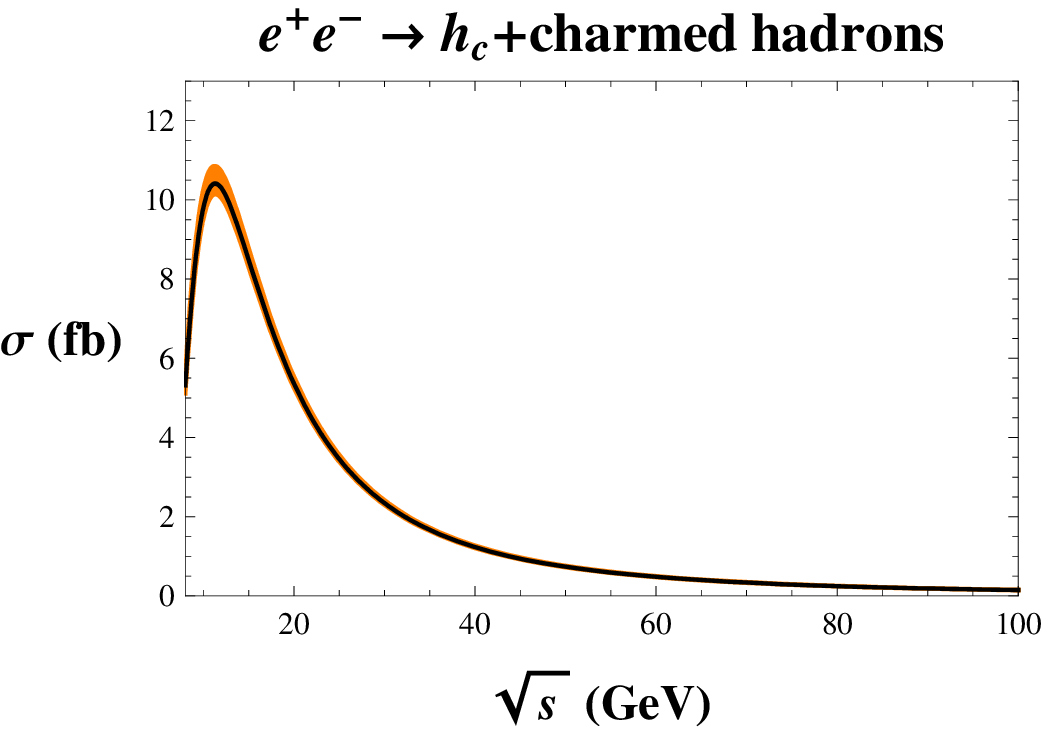}
\caption{The integrated $h_c$ production cross section as a function
of $\sqrt{s}$ for the processes $e^+e^-\to h_c+X_{\rm LH}$ (left
panel) and $e^+e^-\to h_c+X_{c\bar{c}}$ (right panel). The band in
each plot measures the uncertainty brought in by varying the
color-octet matrix element
$\langle\mathcal{O}_8^{h_c}(^1S_0)\rangle$ from 0.0085 to 0.039
${\rm GeV}^3$, where the central curve corresponds to fixing
$\langle\mathcal{O}_8^{h_c}(^1S_0)\rangle$ at $0.02\;{\rm GeV}^3$.
\label{hc:against:sqrts:plot}}
\end{center}
\end{figure}
%--------------------------

Finally, in Fig.~\ref{hc:against:sqrts:plot} we show how the
integrated $h_c$ production rates vary with the center-of-mass
energy. As $\sqrt{s}$ increases, the total cross section of
$e^+e^-\to h_c+X_{\rm LH}$ descends much steeper than that of
$e^+e^-\to h_c+X_{c\bar{c}}$. This can be easily attributed to the
fact that at large $s$, the production rate in the former process
scales as $1/s^2$, while that in the latter case scales only as
$1/s$, as dictated by the fragmentation mechanism.

\subsection{Observation prospects of inclusive $\bm{h}_{\bm c}$ production at $\bm{B}$ factories}

Thus far, there is yet no any experiment at $B$ factories dedicated
to measure the inclusive $h_c$ production rate. Perhaps it is partly
due to the lack of phenomenological incentive, and more importantly,
due to the difficulty of reconstructing the $h_c$ signals.

Our prediction in (\ref{predictions:total:hc:cross:sec:two:chan}),
indicates that a large number of $h_c+$ light hadrons events should
have already been produced at $B$ factories, given the large
integrated luminosity that has been accumulated by two $B$ factory
experiments near the $\Upsilon(4S)$ resonance. As we have shown
before, the prospective measurements of this channel would provide a
promising window to test the color-octet mechanism in quarkonium
production, and impose some useful constraint on the color-octet
matrix element $\langle {\cal O}^{h_c}_8({}^1S_0)\rangle$. In view
of this, there should be sufficient phenomenological impetus for
experimentalists to pursue the measurements.

We can examine the observation potential for inclusive $h_c$
production more quantitatively. Up to present, \textsc{Belle}
experiment has accumulated about $1000\;{\rm fb}^{-1}$ data near the
$\Upsilon(4S)$ resonance. Taking $\sigma[e^+e^-\to h_c+X_{\rm
LH}]\approx 100-400$ fb from
(\ref{predictions:total:hc:crossX:chan:LH}), we thus estimate that
roughly $(1-4)\times 10^5$ $h_c$ events have been produced at
\textsc{Belle} at $\sqrt{s}=10.58$ GeV.

The two known decay channels of the $h_c$ meson are $h_c\to 2(\pi^+
\pi^-)\pi^0$ and $h_c\to \eta_c\gamma$. The multi-pion decay seems
to be a clean and potentially useful tagging mode fr the $h_c$
signal. Nevertheless, due to the greater branching fraction of the
latter channel, it might be of some advantages by utilizing the $E1$
transition $h_c\to \eta_c\gamma$ to reconstruct the $h_c$ signal,
i.e., first reconstruct an $\eta_c$, then check if there exists a
narrow peak around 3525 MeV in the $\gamma+\eta_c$ invariant mass
distribution.

The main experimental difficulty hinges on how to efficiently
reconstruct the $\eta_c$. It seems to be a standard practice for the
\textsc{Belle} Collaboration to reconstruct the $\eta_c$ meson via
several hadronic decay modes, {\it e.g.} $K_S K^+\pi^- + c.c.$,
$\pi^+\pi^-K^+K^-$, $2(K^+K^-)$, $2(\pi^+\pi^-)$, $3(\pi^+\pi^-)$ by
looking for the charged tracks (In fact, they have employed this
technique in searching for the radiative decay processes
$\Upsilon(1S,2S)\to \gamma \eta_c$~\cite{Shen:2010iu,Wang:2011qm}).
A conservative estimate gives that about the $1\%$ $\eta_c$ events
can be reconstructed~\footnote{This estimate is obtained from the
product of detection efficiency and the branching fractions of
aforementioned $\eta_c$ decay channels. With the efficiency taken
with around 25\%, it is reasonable to assume that 1\% $\eta_c$
events will be reconstructed. We thank C.~Z.~Yuan for providing this
estimate and for explaining to us some experimental details.}.

Taking ${\cal B}(h_c\to \gamma \eta_c)\approx 50
\%$~\cite{Ablikim:2010rc}, we then expect roughly $(1-4)\times
10^5\times 50\%\times 1\% = 500-2000$ reconstructed $h_c$ events.

A large number of reconstructed $h_c$ events seems to indicate a
bright observation prospect, however, one must be alert to the
potentially huge combinatorial background, since there are a lot of
pions, kaons and photons in the events. A careful study of the
background level is crucial from the experimental perspective.

%-------------------------------------------
\section{Summary}
\label{summary}
%-------------------------------------------

In this work, we have studied the inclusive production of the $h_c$
meson associated with the light hadrons and with the charmed hadrons
at the $B$ factories, respectively, in the framework of NRQCD
factorization. We have only considered the lowest-order contribution
in $v$, which involves both the color-singlet ${}^1P_1^{(1)}$
channel and the color-octet ${}^1S_0^{(8)}$ channel.

We have explicitly verified that, for $e^+e^-\to h_c+$ light
hadrons, including the color-octet channel is pivotal in rendering
the IR-finite prediction for the inclusive $h_c$ production rate.
Moreover, at $B$ factory energy $\sqrt{s}=10.58$ GeV, within some
reasonable choices for the color-octet matrix element $\langle {\cal
O}^{h_c}_8({}^1S_0)\rangle$, we find that the total $h_c$ production
rate varies in the range between 0.08 and 0.4 pb. Furthermore, this
process is found to be dominated by the color-octet channel. Future
measurement of this process at $B$ factories may provide an explicit
test on the color-octet mechanism, as well as put a useful
constraint on the value of the color-octet matrix element $\langle
{\cal O}^{h_c}_8({}^1S_0)\rangle$.

Since this process is dominated by the color-octet channel, it is
reasonable to expect that the bulk of the $h_c$ events will populate
near the upper end of the momentum spectrum. For the
experimentalists to unambiguously measure the $h_c$ production rate
at the $B$ factories, it is perhaps imperative to first have a
reliable prediction for the momentum spectrum of the
$h_c$~\footnote{We thank C.-P. Shen for stressing this point to us,
and for having performed a detailed Monte Carlo study on
reconstructing the $h_c$ events at the \textsc{Belle} experiment.}.
This direction is certainly worth further investigations.

For the process $e^+e^-\to h_c+$ charmed hadrons, we find that the
total production rate is almost saturated by the color-singlet
channel. The predicted total production cross section at
$\sqrt{s}=10.58$ GeV is about 10 fb if the charm quark mass is taken
to be 1.5 GeV. This is about one order-of-magnitude smaller than the
cross section for $e^+e^-\to h_c+$ light hadrons. We note that this
prediction bears large uncertainty, for it is quite sensitive to the
charm quark mass. If allowing $m_c$ to float from $1.8$ GeV to $1.3$
GeV, this cross section would vary from a few fb to 30 fb. It is
interesting to see whether the future experiments can observe this
process or not.

{\it Note added.}  After this paper was submitted, there has
recently appeared a related work in arXiv~\cite{Wang:2012tz}, which
also studied the inclusive $h_c$ production in $e^+e^-$ annihilation
in NRQCD factorization framework. For the $h_c$ production
associated with the light hadrons, these authors adopted a
factorization scheme different from the ${\rm \overline{MS}}$
scheme, and calculated the integrated NRQCD short-distance
coefficients using some numerical recipe.

%--------------------------------------------------------------------
\begin{acknowledgments}
%--------------------------------------------------------------------
We are grateful to Cheng-Ping Shen and Chang-Zheng Yuan for many
useful discussions on the experimental issues.
%--------------------------------------------------------------------
This research was supported in part by the National Natural Science
Foundation of China under Grant No.~10875130, No.~10935012, DFG and
NSFC (CRC 110), and China Postdoctoral Science Foundation.
%--------------------------------------------------------------------
\end{acknowledgments}
%--------------------------------------------------------------------

%--------------------------------------------------------------------
\appendix

%--------------------------------------------------------------------
\section{Isolating the infrared divergence for $e^+e^-\to c\bar{c}({}^1P_1^{(1)})+gg$
 in dimensional regularization
\label{appendix:DR:isolate:IR:div}%
}

In this Appendix, we explain how to calculate the IR-divergent
integral in (\ref{dsig:div:dz:def}) in dimensional regularization.
The $D$-dimensional 3-body phase space measure $d\Phi_3$ and the
integrand $I_{\rm div}(x_1,z)$ have already been presented in
(\ref{3-body:phase:space:D:dimensions}) and (\ref{Idiv:def}),
respectively. Inspecting these expressions, one can readily
recognize that the IR singularity would arise from two distinct
phase space corners: $z\to 1+r, x_1\to 0$ and $z\to 1+r, x_1\to
1-r\,(x_2\to 0)$, by integrating the first and second terms in
(\ref{Idiv:def}), respectively.

First note that both the integrand and the phase space measure in
(\ref{dsig:div:dz:def}) are symmetric under the interchange
$x_1\leftrightarrow x_2$. A useful shortcut to perform the
$x_1$-integral is to integrate over only the lower half of its
allowed range, then multiply the result by 2:
%--------------------------
\bqa
%--------------------------
& & \int_{a(z)-b(z)}^{a(z)+b(z)} dx_1  = 2\int_{a(z)-b(z)}^{a(z)}
dx_1,
%--------------------------
\label{}
%--------------------------
\eqa
%--------------------------
where $a(z)$ and $b(z)$ are defined in (\ref{def:a(z):b(z)}). As a
simplification, the IR singularity now is solely caused by the first
term in (\ref{Idiv:def}), when integrated over the phase space
corner $z \to 1+r, x_1\to 0$. The integral involving the second term
now becomes IR finite, thereby can be worked ont directly in $4$
dimensions.

We then face the IR-divergent integral of the following type:
%--------------------------
\bqa
%--------------------------
I(z) &=& \int^{a(z)}_{a(z)-b(z)} dx_1 {g(x_1,z)\over (1+r-z-x_1)^2
x_1^{2\epsilon}(1-\cos^2 \theta)^\epsilon},
%--------------------------
\label{I(z):def:integral:over:x1}
%--------------------------
\eqa
%--------------------------
where $\cos\theta$ as a function of $z$ and $x_1$ has been given in
(\ref{theta:def}), and $g(x_1,z)$ is an arbitrary function that is
regular at $z=1+r$. Note that the $\epsilon$-dependent factors in
the integrand come from the $D$-dimensional 3-body phase space
measure, playing the role of the IR regulator.

A brute-force calculation of $I(z)$ while keeping the full
$\epsilon$ dependence is a challenging task, if not impossible. It
is desirable if one can make the IR divergence explicit prior to
carrying out the $x_1$ integration. This is indeed feasible, and in
the analysis of the decay process $\chi_{bJ}\to c\bar{c}g$, Bodwin
{\it et al.}~\cite{Bodwin:2007zf} have elaborated on how to fulfill
such a goal by utilizing a simple trick. In below, we will employ a
strategy that is closely analogous to theirs~\footnote{We note that
the process $\eta_b\ \to\chi_{cJ}+gg$ ($J=0,1,2$) has recently been
analyzed in NRQCD factorization~\cite{He:2009sm}, which shares
essentially the identical kinematics as our process. The methodology
of isolating the IR singularity in DR in \cite{He:2009sm} is similar
to what is adopted in \cite{Bodwin:2007zf}.}.

The key is to realize that $x_1^+ x_1^-=a^2(z)-b^2(z)=1+r-z$. As
suggested by this identity, it may seem advantageous to introduce a
new integration variable $t$:
%--------------------------
\bqa
%--------------------------
t &=&  {1+r-z\over x_1}=a(z)+b(z)\cos\theta.
%--------------------------
\label{new:variable:t}
%--------------------------
\eqa
%--------------------------
The integration range for this new variable turns out to be $t_0
\leq t \leq a(z)+b(z)$, where
%--------------------------
\bqa
%--------------------------
t_0 &=&  {2(1+r-z) \over 2-z}.
%--------------------------
\eqa
%--------------------------

In term of the new variable $t$, equation
(\ref{I(z):def:integral:over:x1}) becomes
%--------------------------
\bqa
%--------------------------
I(z) &=&  {1\over (1+r-z)^{1+2\epsilon}} \int^{a(z)+b(z)}_{t_0} \!
dt \, {t^{2\epsilon}\over (1-t)^2} {g(t,z)\over (1-\cos^2
\theta)^\epsilon}.
%--------------------------
\label{I(z):def:integral:over:t}
%--------------------------
\eqa
%--------------------------
In light of (\ref{new:variable:t}), $\cos\theta$ now should be
understood as a function of $t$, $a(z)$ and $b(z)$.

With the aid of the familiar identity about distributions, we can
express $(1+r-z)^{-1-2\epsilon}$ as
%--------------------------
\bqa
%--------------------------
{1 \over (1+r-z)^{1+2\epsilon}} &=& - {\delta(1+r-z)\over 2\epsilon
\,(1- \sqrt{r})^{4\epsilon}} + \bigg[{1 \over 1+r-z} \bigg]_+ -
2\epsilon \bigg[{\ln(1+r-z) \over 1+r-z }  \bigg]_+ \cdots.
%--------------------------
\label{identity:distributions}
%--------------------------
\eqa
%--------------------------
As advertised, we have successfully separated the IR pole that is
accompanied with the $\delta(1+r-z)$ function. The IR-regular
remainders are partially encoded in the ``+"-functions, whose
integration property has been specified in
(\ref{plus:function:convolution}). To the desired accuracy, the
logarithmic ``+" function is not needed in this work.

One can be readily convinced that the integral in
(\ref{I(z):def:integral:over:t}) is finite in the limit $\epsilon\to
0$. Therefore, to our intended accuracy, it can be evaluated by
simply Taylor-expanding the integrand through the first order in
$\epsilon$.

Using the master formulas (\ref{I(z):def:integral:over:t}) and
(\ref{identity:distributions}), it is now a straightforward exercise
to reproduce the analytic expression of $d \hat{\sigma}_{\rm
div}/dz$ as given in (\ref{dsig:div:dz:expression}). One can further
integrate this expression over $z$ to arrive at
%--------------------------
\bqa{\label{IR cross section}}
%--------------------------
\hat{\sigma}_{{\rm div}}&=& {128\pi e_c^2 \alpha^2 C_F \alpha_s^2
(1-r) \over 3 m_c^2 s}\frac{c_\epsilon(4\pi)^\epsilon}
{\Gamma(1-\epsilon)} \bigg[-\frac{1}{\epsilon_{\rm IR}}-2\ln
\frac{\mu^2}{4m_c^2}
%--------------------------
\nn\\
%--------------------------
& & + 4\ln(1-r) - {1-3r \over 1-r}\ln r-1
  \bigg].
%--------------------------
%--------------------------
\eqa
%--------------------------

\section{$\bm{\eta}_{\bm c}$ production associated with charmed hadrons in
${\bm e}^{\bm +}{\bm e}^{\bm -}$ annihilation}
\label{appendix:etac:c:cb:dis:frag}%

In this Appendix, we consider $e^+e^-\rightarrow \eta_c+$ charmed
hadrons at lowest order in $v$ and $\alpha_s$ (note that
$e^+e^-\rightarrow \eta_c+gg$ is forbidden by the charge conjugation
invariance). The result presented in this section can be viewed as a
byproduct of the analysis made in Sec.~\ref{matching:hc:ccbar}. We
note that, the process $e^+e^-\rightarrow \eta_c+c\bar{c}$ has
already been investigated some time ago~\cite{Liu:2003jj}, and our
calculation may serve as an independent check.

At the lowest order in $v$, the NRQCD factorization formula for the
$\eta_c$ inclusive production reads~\cite{Bodwin:1994jh}:
%--------------------------
\bqa
%--------------------------
d\sigma[e^+e^-\rightarrow \eta_c+ X_{c\bar{c}}] = {dF^{\rm Charm}_1
\over m_c^2} \langle \mathcal{O}_1^{\eta_c}(^1S_0) \rangle +
O(\sigma v^2),
%--------------------------
\label{NRQCD:factorization:formula:etac}
%--------------------------
\eqa
%--------------------------
where the color-singlet production operator
$\mathcal{O}_1^{\eta_c}(^1S_0)$ has been defined in
\cite{Bodwin:1994jh}, $dF^{\rm Charm}_1$ is the corresponding
short-distance coefficient. Unlike the $P$-wave charmonium
production, here we are justified to ignore the color-octet channel
since its relative importance is suppressed by $v^4$.

The perturbative matching calculation for the color-singlet
short-distance coefficient is completely analogous to
Sec.~\ref{matching:hc:ccbar}, and we directly present
the result:
%--------------------------
\bqa
%--------------------------
&&{dF^{\rm Charm}_1 \over dz }  =\frac{32\pi
e_c^2\alpha^2\alpha_s^2}{81m_c s(2-z)^2 z^3} \Bigg\{ \frac{2 z }{3
(2-z)^4}\sqrt{\frac{(1-z) (z^2-4 r)}{1+r-z}}\times \nn\\
%--------------------------
   &&\bigg[ 96 r^3 (1+r) -96
   r^2 (4+6 r+r^2) z  +16
   (6-25 r+38 r^2+43 r^3-2 r^4
   ) z^2 \nn\\
%--------------------------
   && -8
   (24-138 r+58 r^2+51 r^3-r^4) z^3 +(112-944 r+372 r^2+118 r^3-6
   r^4
) z^4\nn\\
%--------------------------
   &&-2
   (24-120 r+62 r^2+r^3) z^5+(54+5 r+13 r^2) z^6-(28+11 r) z^7
  +6
   z^8
   \bigg]\nn\\
&& + r \ln \bigg(\frac{z\sqrt{1+r-z}+\sqrt{(1-z) (z^2-4
   r)}}{z\sqrt{1+r-z}-\sqrt{(1-z) (z^2-4
   r)}}\bigg)
    \bigg[8 r^3-32 r^2 z-2
   (4-6 r-r^3) z^2\nn\\
%--------------------------
&&-4 r (1+r) z^3+(10+r) z^4
   \bigg]
\Bigg\}.
%--------------------------
\label{dF1:charm:dz:eta_c:c:cbar}
%--------------------------
\eqa
%--------------------------
Our expression seems to be considerably compact than its counterpart
in \cite{Liu:2003jj}. We have numerically checked that once using
their phenomenological input parameters, we can reproduce their
predicted production rate for $\eta_c+X_{c\bar{c}}$ at $B$ factory
energy.

Similar to the fragmentation function of $c$ into $h_c$ in
(\ref{Fragmentation:function:NRQCD:factorization}), the
fragmentation function of $c$ into $\eta_c$ can also be factorized
as
%--------------------------
\bqa
%--------------------------
D_{c\to \eta_c}(z) =   d_1^{\eta_c}(z) {\langle
\mathcal{O}_1^{\eta_c}({}^1S_0) \rangle \over m_c^3}+ O(v^2),
%--------------------------
\label{Frag:function:NRQCD:factorization:etac}
%--------------------------
\eqa
%--------------------------
where $d_1^{\eta_c}(z)$ is the color-singlet coefficient function,
which agrees with the one give in \cite{Braaten:1993mp}.

Following the steps described in
Sec.~\ref{Frag:Func:c:hc:int:X:section}, it is straightforward to
identify $d_1^{\eta_c}(z)$ by taking the asymptotic limit to
(\ref{dF1:charm:dz:eta_c:c:cbar}):
%--------------------------
\bqa
%--------------------------
 d_1^{\eta_c}(z)&=&{m_c\over 2 \check{\sigma}} {d F_1^{\rm Charm}
\over dz}\bigg|_{\rm asym} =  {16 \alpha_s^2 z (1-z)^2 \left(48+8
z^2-8 z^3+ 3 z^4 \right)\over 243(2-z)^6},
%--------------------------
\eqa
%--------------------------
which fully agrees exactly with the fragmentation function of $c\to
\eta_c$ first obtained in \cite{Chen:1993ii,Braaten:1993mp}.
Obviously, this expression only differs with $d_8^{h_c}$ in
(\ref{d1:d8:short-dist:coef:frag:function}) by a color factor.

Finally we assess the integrated cross section for
$e^+e^-\rightarrow \eta_c+ X_{c\bar{c}}$. It is difficult to obtain
the analytic result by integrating (\ref{dF1:charm:dz:eta_c:c:cbar})
over $z$. Nevertheless we are content with knowing its asymptotic
behavior by resorting to fragmentation approximation:
%--------------------------
\bqa
%--------------------------
& & \sigma[e^+e^-\to \eta_c+ X_{c\bar{c}}]\bigg|_{\rm asym}=
2\check{\sigma} \int^1_0\!dz\,D_{c\to \eta_c}(z)
%--------------------------
\nn\\
%--------------------------
%--------------------------
& & = 16  \alpha_s^2 \check{\sigma } {773-1110\ln2 \over 1215}
{\langle {\mathcal O}^{\eta_c}_1(^1S_0) \rangle \over m_c^3}.
%--------------------------
\eqa
%--------------------------
This expression is compatible with the one give in
\cite{Braaten:1993mp}.

%%%%%%%%%%%%%%%%%%%%%%%%%%%%%%%%%%%%%%%%%%%%%%%%%%%%%%%%%%%%%%%%%%%%%%%%%%%%%%

\end{document}